\documentclass[
reprint,
showpacs,
preprintnumbers,
bibnotes,
amsmath,amssymb,
aps,
prd,
]{revtex4-1}

\usepackage{graphicx}
\usepackage{dcolumn}
\usepackage{bm}
\usepackage{color}
\usepackage{multirow}
\usepackage{mathrsfs}
\usepackage{amssymb}

\begin{document}


\title{Framework for maximum likelihood analysis 
of neutron $\bm{\beta}$ decay observables \\
to resolve the limits of the $\bm{V-A}$ law}

\author{S.\ Gardner}
\email{\texttt{gardner@pa.uky.edu}}
\author{B.\ Plaster}
\email{\texttt{plaster@pa.uky.edu}}
\affiliation{Department of Physics and Astronomy, University of Kentucky,
Lexington, Kentucky 40506-0055 USA}


\begin{abstract}
We assess the ability of future neutron $\beta$ decay measurements of
up to ${\cal O}(10^{-4})$ precision 
to falsify the standard model, particularly the $V-A$ law, 
and to identify the dynamics beyond it. To do this, 
we employ a maximum likelihood statistical framework which incorporates 
both experimental and theoretical uncertainties. 
Using illustrative combined global 
fits to Monte Carlo pseudodata, we also quantify 
the importance of experimental 
measurements of the energy dependence of the angular correlation
coefficients as input to such efforts, 
and we determine the 
precision to which ill-known ``second-class'' hadronic matrix elements
must be determined in order to exact such tests. 
\end{abstract}


\maketitle

\section{Introduction}
\label{sec:introduction}

Inspired by the pioneering global fits of the
Cabibbo-Kobayashi-Maskawa (CKM) matrix developed by the 
CKMfitter Group \cite{hocker01,charles05} and the UTfit
Collaboration \cite{ciuchini00,ciuchini01,Bona:2005vz,Bona:2006ah,Bona:2007vi}
for the interpretation of flavor-physics 
results from the $B$ factories and the 
Tevatron, we outline the prospects for the elucidation of physics beyond the 
standard model (BSM) via 
a global fit of neutron $\beta$ decay observables, including the 
lifetime and the energy-dependence of the angular correlation
coefficients.  Our 
global fit, which we term $n$Fitter, employs 
a maximum likelihood statistical framework which accounts for both 
experimental and theoretical uncertainties, with the latter arising
primarily from the poorly known weak hadronic second-class currents.

The elucidation of a ``$V-A$'' law \cite{Feynman:1958ty,Sudarshan:1958vf}
in the mediation of 
low-energy weak interactions played a crucial role in the rise
of the standard model (SM)~\cite{Hoddeson:1997hk}. 
A variety of well-motivated SM extensions 
speak to the possibility of new dynamics at the Fermi scale, which we identify
via 
$v= (2\sqrt{2} G_F)^{-1/2} \approx 174\,{\rm GeV}$,  
and, concomitantly, to tangible 
departures 
from the $V-A$ law in low-energy weak processes, be it through 
scalar or tensor interactions, or right-handed currents. 
Alternatively, e.g., 
new, light degrees of freedom could appear 
and 
yield violations of the $V-A$ law --- if probed 
at sufficient 
experimental resolution. 
At the same time, an ongoing vigorous experimental program for 
precision measurements of neutron $\beta$ decay observables with cold 
and ultracold neutrons
\cite{nico05,severijns06,abele08,nico09,dubbers11,severijns11,few11} 
exists with an overarching goal of realizing 
bettered assessments of the limits 
of the SM.  The experimental 
effort focuses on measurements of two general types of
observables: angular correlation
coefficients, which parametrize the angular correlations between the
momenta of the various decay products and/or 
the spin of the initial state neutron or electron  
in the differential decay rate, and the neutron lifetime $\tau$. 
Of the future experimental plans reviewed in 
Refs.~\cite{nico05,severijns06,abele08,nico09,dubbers11,severijns11,few11,Cirigliano:2013xha,martin13}, the claim of ultimate precision rests with the PERC 
experiment \cite{Dubbers:2007st}, for a sensitivity of up to $10^{-4}$ precision.

The angular correlation coefficients 
$a$, $b$, $A$, $B$, and $D$~\cite{jackson57}, of which
only $a$, $A$, and $B$ are measured to be nonzero~\cite{PDG}, 
describe the distribution in electron and neutrino directions and
electron energy in neutron decay. 
Since the nucleon mass is 
markedly larger than the neutron-proton mass difference, 
a recoil expansion of the differential decay rate of the
neutron in terms of 
the ratio of various small energy scales 
to the nucleon mass 
is extremely efficient. 
The $a$, $A$, and $B$ correlation coefficients in this order, 
which we denote with a ``$0$'' subscript, are 
functions only of $\lambda \equiv g_A/g_V >0$ in the SM, where $g_A$ and $g_V$ are the
weak axial-vector and vector coupling constants; this is 
tantamount to 
the $V-A$ law.
Note that for consistency with earlier work
\cite{bender68,holstein74,gardner01},
we employ a $\lambda > 0$ sign convention, so that the 
sign of the $\gamma_5$ terms in the weak currents are 
chosen opposite to that of the Particle
Data Group \cite{PDG}.  We also assume that $\lambda$ is real, which 
has been  established beyond our assumed level of 
sensitivity in the observables we consider~\cite{mumm11,chupp12}. 
Thus, extractions of $a_0$, $A_0$, and $B_0$ from the measured parameters 
$a$, $A$, and $B$ determine $\lambda$ in the SM, which, by itself, is
a fundamental parameter of the weak interaction.
Using the measured values for 
$\tau$ and $\lambda$ determines $g_V$
and $g_A$ independently. The former, 
once radiative corrections are applied~\cite{czarnecki04,marciano06}, 
yields the CKM matrix element $V_{ud}$.
A neutron-based value for $V_{ud}$ \cite{PDG} is not yet competitive
with the result 
extracted from measurements of the 
$ft$ values in superallowed $0^+ \rightarrow 0^+$ nuclear
$\beta$ decays \cite{hardy09,towner10}. 
Such efforts are nevertheless well-motivated in 
that they are
not subject to the nuclear structure corrections which must be applied
in the analysis of $0^+ \rightarrow 0^+$ transitions. 
We set such prospects aside and 
focus on the possibility of discovering dynamics beyond the SM
through an assessed quantitative violation of the $V-A$ law. 

We consider neutron beta decay observables exclusively, 
though the framework --- and fits --- 
we employ can readily be enlarged to encompass all 
low-energy semileptonic processes involving 
first-generation quarks. This is possible because 
we can employ, rather generally,  
a single, quark-level effective Lagrangian, 
at leading power in $v/\Lambda$~\cite{Buchmuller:1985jz,Grzadkowski:2010es}, 
for all such processes~\cite{Cirigliano:2009wk,Cirigliano:2012ab}, 
where we refer to Ref.~\cite{Cirigliano:2013xha} 
for a review. To realize this we need only assume that 
new physics
appears at an energy scale $\Lambda$ in excess of the scale $v$,
crudely that of the $W$ and $Z$ masses. This allows the construction of
an effective Lagrangian in terms of operators of mass dimension $d$ 
with $d>4$ \cite{Appelquist:1974tg}, 
where the 
nonobservation of non-SM invariant operators in 
flavor physics~\cite{Isidori:2010kg,Bona:2007vi,charles05}, 
allows us to impose SM electroweak gauge invariance 
in its construction. At leading power in $v/\Lambda$, 
there are then precisely ten dimension-six operators describing the 
semileptonic decay of the $d$ quark. The coefficients of 
these operators can be determined, or at least constrained, through
a global fit of $\beta$ decay observables --- in neutrons and nuclei --- 
and of meson decay observables as well. 
This quark-level description, upon 
matching to a
nucleon-level effective theory~\cite{bhattacharya12}, yields a
one-to-one map to the terms of the effective Hamiltonian constructed
by Lee and Yang \cite{leeyang56}, admitted by Lorentz invariance and
the possibility of parity violation \cite{leeyang56}.  The latter
framework is employed in the usual analysis of the angular
correlations in beta decay~\cite{jackson57}. The 
construction of the underlying quark-level effective theory 
allows the inclusion of meson observables on the same footing. 

Global fits of the Lee-Yang coefficients have been made to beta decay
observables, particularly of the phase-space-integrated angular
correlation coefficients~\cite{severijns06,konrad10}.  As we 
develop explicitly, the advantage of the current approach is that we
can take theory errors into direct account in the
optimization procedure. 
We focus on the simultaneous fit 
of the angular correlation coefficients $a$ and $A$ in neutron decay and of 
the neutron lifetime to limit the appearance of tensor and scalar
interactions from physics BSM, in part because the 
theory errors in this case are spanned by the ill-known nucleon matrix elements of
particular local operators, which 
presumably and eventually can be computed in QCD using the 
techniques of lattice gauge theory.

The empirically 
determined correlation coefficients
and neutron lifetime, 
fixed to some precision, can be used to set limits on new dynamics. 
In this regard it is natural to focus on the 
Fierz interference term $b$ because it is
linear in new physics couplings. That is, 
it is possible to discover evidence of scalar or tensor interactions, 
through a measurement of $b$ 
in excess of 
the $\sim 10^{-3}$ level expected in the SM from recoil-order
effects \cite{holstein74}. To date there have been no 
published results for $b$ in neutron $\beta$ decay, although several
efforts are underway. Searches for scalar and tensor interactions
have also been pursued via measurements of the decay electron's
transverse polarizations with respect to the neutron spin \cite{kozela09,kozela12},
though no new experiments of such ilk are currently planned. 
Information on $b$ can be gleaned in different ways. 
Its presence modifies the shape of the electron
energy spectrum in the differential decay rate, and direct searches
probe that. It is also subsumed in measurements of the energy dependence
of the $a$, $A$, and $B$ correlation coefficients \cite{gluck95,bhattacharya12}, 
so that we wish to consider the additional 
experimental observables offered by (electron) energy dependence with some care. 
The latter pathways offer ``indirect'' access to $b$ 
and require greater
statistical control than spectral shape measurements for fixed
sensitivity to scalar and tensor interactions. However, they are also
less sensitive to systematic errors, both experimental and
theoretical. In this first paper we focus on access to scalar and
tensor interactions through measurements of $a$, $A$, and $\tau$, as
we now explain.

The $V-A$ description of neutron
$\beta$ decay includes contributions from six possible weak hadronic
currents which give rise, at recoil order, to 
energy-dependent expressions for the angular correlation coefficients.
However, the most recent extractions of $\lambda$ from $A$
\cite{abele02,mund13,liu10,plaster12,mendenhall13}, 
which yield, at present, 
the most precise determination of $\lambda$, have 
assumed the validity of the conserved-vector-current (CVC) 
hypothesis (i.e., the CVC value for
the weak magnetism coupling) and neglected second-class currents.
This is certainly reasonable at the current level of
experimental precision, although measurements of $A$ at the $\sim
0.5$\% level of precision are, in principle, sensitive to the 
weak magnetism coupling, which contributes to the asymmetry at the
$\sim 1.5$\% level.  With the anticipated increased sensitivity to
$a$, $A$, and $B$ in next-generation experiments, we note
that precise measurements of
the energy dependence of the $a$ and $A$ correlations 
offer the possibility to test CVC 
and to search for second-class currents independently \cite{gardner01}. 
In  $n$Fitter
we fit the energy dependence of the
angular correlation coefficients directly and thus need not assume either 
the validity of the CVC hypothesis or 
the absence of 
second-class currents, yielding a framework for a 
robust test of the validity of the $V-A$
description of neutron $\beta$ decay.

The outline of the remainder of this paper is as follows.  First, we
review the formalism for the analysis of neutron $\beta$ decay 
observables 
in Sec.\ \ref{sec:formalism}.  We then 
briefly review in Sec.\ \ref{sec:uncertainties} the experimental
status of, or limits on, the various SM 
parameters relevant to our global fit, such as second-class currents.
We then describe the maximum likelihood approach to our global fit in
Sec.\ \ref{sec:likelihood}, where we discuss the construction of our
likelihood function, as well as the inclusion of experimental and theoretical
uncertainties.
We illustrate the prospects of our global
fit with a few numerical examples employing the 
frequentist, or specifically $R$fit \cite{hocker01,charles05}, 
statistical procedure,
reserving the use of alternative statistical procedures for later work.
We then show examples of the 
results from $n$Fitter fits to Monte Carlo generated pseudodata
in Sec.\ \ref{sec:example_fits} under various scenarios, and we
quantify via these examples the statistical impact that future
improvements in the precision of neutron $\beta$ decay observables
should have on the assessment of the validity of the SM.  We
also discuss the extent to which theoretical uncertainties in the presently 
poorly known second class contributions limit such 
assessments, and we can extract from such studies the precision to 
which they should be established to obviate that impact.  
Finally, we conclude with a brief summary in Sec.\
\ref{sec:summary}.

\section{Formalism for Neutron $\bm{\beta}$-Decay Observables}
\label{sec:formalism}

If we suppose the low-energy, effective 
weak interaction, employing explicit $n$ and $p$ 
degrees of freedom, is mediated by the ten dimension-six operators 
enumerated by Lee and Yang \cite{leeyang56}, 
then the differential decay rate for neutron $\beta$ 
decay takes the form \cite{jackson57}
\begin{eqnarray}
&&\displaystyle{\frac{d\Gamma}{dE_e d\Omega_e d\Omega_\nu}}
= \frac{1}{(2\pi)^5} p_e E_e (E_0 - E_e)^2 \xi 
\nonumber \\
&&~~\times \left[1 + b\frac{m_e}{E_e} +
a\frac{\vec{p}_e \cdot \vec{p}_\nu}{E_e E_\nu} \right. \nonumber \\
&&~~~~~~~\left. +~\langle \vec{\sigma}_n \rangle \cdot
\left( A\frac{\vec{p}_e}{E_e} +
B\frac{\vec{p}_\nu}{E_\nu} +
D\frac{\vec{p}_e \times \vec{p}_\nu}{E_e E_\nu}\right) \right]\,,
\label{eq:W_angular_distribution}
\end{eqnarray}
where we refer to Ref.~\cite{jackson57} for the explicit form of $\xi$ and
the correlation coefficients 
in terms of the parameters of the Lee-Yang Hamiltonian \cite{leeyang56},
noting Ref.~\cite{Cirigliano:2013xha} 
for a discussion of the connection to modern conventions. 
We use $E_e$ ($E_\nu$) and $\vec{p}_e$ ($\vec{p}_\nu$) to denote, 
respectively, the electron's (antineutrino's) total energy and
momentum, where $E_0$
is the electron endpoint energy, and $\langle \vec{\sigma}_n \rangle$ is the
neutron polarization.  

The
Coulomb corrections to Eq.~(\ref{eq:W_angular_distribution})
are also known \cite{jackson57npa} and modify
the expression most notably in terms of a multiplicative 
Fermi function $F(Z,E_e)$ \cite{fermi34}. The phase-space integrated 
Fermi function and corrections to it
have been studied in great detail \cite{wilkinson82,wilkinson98}; 
we omit it, as well as the outer radiative correction~\cite{Sirlin:1967zza}, 
 in the generation of the Monte Carlo pseudodata for 
our decay correlation studies as we are interested 
in $a(E_e)$ and $A(E_e)$, which are accessed 
through asymmetry measurements for which such effects 
only lead to a slight modification of the 
relative statistics (via the spectral shape).

The $D$ term is a naively time-reversal-odd observable: 
 a value for $D$ in excess of the $\sim 10^{-5}$ level attributed to 
SM final-state interaction effects 
\cite{callan67,ando09} would reveal the existence of new CP-violating 
interactions at the Lagrangian level (assuming CPT holds). 
The current level of experimental precision places stringent 
constraints on any such new effects \cite{mumm11,chupp12}.

In what follows we report expressions for the correlation coefficients
which include the tree-level new physics of the Lee-Yang Hamiltonian
and the contributions of the usual $V-A$ terms through recoil
order. In realizing this the strong interaction plays an essential
role: the matrix elements of the vector $V$ and axial-vector $A$
currents are described by six distinct form factors.  We find it
immensely useful to note the quark-level effective theory which
underlies the Lee-Yang couplings
\cite{Cirigliano:2009wk,bhattacharya12,Cirigliano:2012ab}; it makes clear
the separation of the QCD physics which underlies the hadronic matrix
element calculation from the nominally higher-energy physics encoded
in the effective low-energy constants.  As per
Refs.~\cite{bhattacharya12,Cirigliano:2012ab} we map the Lee-Yang
effective couplings $C_i, C_i^\prime$ with $i\in \{ V ,A, S, T \}$ to
$C_i^{(\prime)}\equiv (G_F/\sqrt{2})V_{ud} \tilde C_i^{(\prime)}$ and
note the hadronic matrix elements needed in $\beta$ decay
are~\cite{weinberg58}
\begin{widetext}
\begin{eqnarray}
\langle p(p') | \bar u \gamma^\mu d | n (p) \rangle &\equiv& \overline{u}_p (p')
\left[ f_1(q^2)\gamma^\mu - i \frac{f_2(q^2)}{M}\sigma^{\mu\nu}q_\nu +
\frac{f_3(q^2)}{M}q^\mu \right]
u_n(p) \,, 
\label{eq:form_factorV} 
\\
\langle p(p') | \bar u \gamma^\mu \gamma_5 d | n (p) \rangle &\equiv& \overline{u}_p (p') \left[
g_1(q^2)\gamma^\mu\gamma_5 - i\frac{g_2(q^2)}{M}\sigma^{\mu\nu}
\gamma_5 q_\nu + \frac{g_3(q^2)}{M}\gamma_5 q^\mu  \right]u_n(p) \,, 
\label{eq:form_factorA}
\\
\langle p(p') | \bar u  d | n (p) \rangle &\equiv& \overline{u}_p (p') g_S(q^2) u_n(p) \,,
\label{eq:form_factorS}
\\
\langle p(p') | \bar u \sigma_{\mu\nu}  d | n (p) \rangle 
&\equiv& \overline{u}_p (p')
\left[ g_T(q^2)\sigma^{\mu\nu}  + g_T^{(1)}(q^2)(q^\mu\gamma^\nu - q^\nu\gamma^\mu) \right.
\nonumber \\
&& \left.  + g_T^{(2)}(q^2)(q^\mu P^\nu - q^\nu P^\mu)   + 
g_T^{(3)}(q^2)(\gamma^\mu \slash{\!\!\!q} \gamma^\nu - \gamma^\nu \slash{\!\!\!q} \gamma^\mu) 
\right]u_n(p) \,,
\label{eq:form_factorT}
\end{eqnarray}
\end{widetext}
where $q\equiv p'-p$ denotes the momentum transfer, $P\equiv p'+p$, 
 and $M$ is the neutron mass. In neutron $\beta$ decay, the $q^2$-dependent terms 
are of next-to-next-to-leading order (NNLO) in the recoil expansion, 
noting $f_1(0)$ and $g_1(0)$ appear in leading order (LO), 
and hence are of 
negligible practical relevance. Consequently, 
we replace, as usual, the form factors with their values at zero momentum transfer. 
We note $f_1(0) \equiv g_V$ is the vector coupling constant given by 
$g_V = 1$ under CVC; 
$f_2(0) \equiv f_2$ is the weak magnetism
coupling constant given by $(\kappa_p - \kappa_n)/2$ under CVC, 
noting $\kappa_{p(n)}$ 
is the anomalous magnetic moment of the proton (neutron); 
$f_3(0) = f_3$ is the induced scalar coupling
constant; $g_1(0) = g_A$ is the axial vector coupling constant;
$g_2(0)=g_2$ is the induced tensor coupling constant; and $g_3(0) =
g_P$ is the induced pseudoscalar coupling constant. The CVC predictions
have SM corrections in NNLO. 
 The contributions of
$f_1$, $f_2$, $g_1$, and $g_3$ to the hadronic current are termed
first-class currents, whereas those of $f_3$ and $g_2$ are termed
second-class currents, due to their transformation properties under
$G$-parity \cite{weinberg58}. The latter quantities, $f_3$ and $g_2$,  
vanish in the SM up to quark mass effects which break flavor symmetry; we discuss
their estimated size in Sec.~\ref{sec:uncertainties}. 

Of particular interest to us are the scalar and tensor interactions, 
as establishing their existence at current experimental limits 
would signify the presence of physics BSM. The 
matching of the quark-level to nucleon-level effective theories 
at LO in the recoil expansion yields: 
\begin{eqnarray}
&& \tilde C_S = g_S (\epsilon_S + \tilde \epsilon_S) \,,\nonumber \\
&& \tilde C_S^\prime = g_S (\epsilon_S - \tilde \epsilon_S) \,,\nonumber \\
&& \tilde C_T = 4g_T (\epsilon_T + \tilde \epsilon_T) \,, \nonumber \\
&& \tilde C_T^\prime = 4 g_T (\epsilon_T - \tilde \epsilon_T) \,,
\label{eq:defST}
\end{eqnarray}
where the $\epsilon$ coefficients are the low-energy constants of the
quark-level effective theory of Refs.~\cite{bhattacharya12,Cirigliano:2012ab}. 
We have neglected the matrix elements $g_T^{(i)}$ 
with $i\in 1,2,3$ in realizing
this expression and thus, for consistency, shall neglect 
the scalar and tensor contribution to recoil order terms
in all that follows. 
Bhattacharya et al. \cite{bhattacharya12} have employed a $R$fit scheme to 
determine the impact of improved lattice estimates of $g_S$ and $g_T$
on the limits on the quark-level low-energy coeffcients for given 
experimental sensitivities to $\tilde C_{S,T}^{(\prime)}$.

In unpolarized neutron $\beta$ decay, the unpolarized differential
distribution 
relevant for a measurement of $a$,
neglecting terms beyond next-to-leading order in the recoil expansion but 
accounting for all six possible form factors, is of the form
\cite{zhang01}
\begin{widetext}
\begin{eqnarray}
\frac{d^3 \Gamma}{dE_e d\Omega_{e\nu}} &\propto&
M^4 R^4 \beta x^2 (1-x)^2 \times \nonumber \\
&& {\Xi} \left[1 + 3 Rx  + Rx
\left( \frac{4\lambda(1 + \kappa_p - \kappa_n)}{1 + 3\lambda^2} \right) -
2 R \left( \frac{\lambda^2 + \lambda + \lambda(\kappa_p - \kappa_n)}
{1 + 3\lambda^2}\right) - 4 R \left( \frac{\lambda g_2}{1 + 3\lambda^2} \right)
\right. \nonumber \\
&& \left.~~~~~~~~  
- \frac{\epsilon}{Rx} \left( 
\frac{1 + 2\lambda + \lambda^2 + 2\lambda(\kappa_p - \kappa_n)} 
{1 + 3\lambda^2} \right) + 
2 \frac{\epsilon}{Rx} \left(\frac{f_3 - \lambda g_2}{1 + 3\lambda^2} 
\right) \right] \times \nonumber \\
&& ~~\left[1 + b_{\mathrm{BSM}}\frac{m_e}{E_e} + 
a_1 \beta \cos\theta_{e\nu} +
a_2 \beta^2 \cos^2 \theta_{e\nu} \right] \,,
\label{eq:parent_a}
\end{eqnarray}
where $\theta_{e\nu}$ is the electron-antineutrino opening angle and 
$\beta \equiv |\vec{p}_e|/E_e$. The structure of this expression 
serves as a {\it de facto} definition 
of $a\equiv a_1+a_2\beta\cos\theta_{e\nu}$ 
and $b_\mathrm{BSM}$ in recoil order. 
It follows that of Ref.~\cite{jackson57} if recoil terms are neglected and is 
that of Ref.~\cite{bilenski60} if $b_{\rm BSM}=0$. 
Note that in writing
the recoil contributions we have neglected terms 
of ${\cal O}(\epsilon_S g_S,\epsilon_T g_T)R$. 
Moreover, 
\begin{eqnarray}
a_1 &=& a_0 + \frac{1}{(1 + 3\lambda^2)^2}
\Big[4\lambda(1 + \lambda + \lambda^2 + \lambda^3 + 2f_2 + 2f_2\lambda^2)R +
(1 + 2\lambda - 2\lambda^3 - \lambda^4 + 4f_2\lambda - 4f_2\lambda^3)
\frac{\epsilon}{Rx} \nonumber \\
&&~~~~~~~~~~~~~~~~~~~~~~~-
\left[8\lambda(1 + 2f_2 + \lambda^2 + 2f_2\lambda^2) +
3(1 + 3\lambda^2)^2\right]Rx \nonumber \\
&&~~~~~~~~~~~~~~~~~~~~~~~+
[2(\lambda - \lambda^3)g_2 + 2(\lambda^2 - 1)f_3]\frac{\epsilon}{Rx}
+ 8\lambda(1 + \lambda^2)g_2 R \Big]\,, 
\label{eq:arecoil}\\
\nonumber \\
a_2 &=& \frac{3(\lambda^2 - 1)}{(1 + 3\lambda^2)} Rx\,,
\end{eqnarray}
with $\lambda = g_A/g_V > 0$ in the SM, 
and the kinematic factors $\epsilon$, $R$, and $x$ are defined
according to
\begin{equation}
\epsilon = \left(\frac{m_e}{M}\right)^2\,,~~~~~
R = \frac{E_0}{M}\,,~~~~~
x = \frac{E_e}{E_0}\,.
\end{equation}
\end{widetext}

\noindent 
The computations of Ref.~\cite{Harrington:1960zz} have been repeated in 
deriving these forms, 
and the results are consistent up to the $f_3$ 
terms \cite{gardner01}. 
They are also consistent with Ref.~\cite{holstein74}, as well as with 
Ref.~\cite{bilenski60}, noting $f_3=g_2=0$ in the latter. 
These comparisons are all within the context of $V-A$ theory. 

We use $R$ itself, noting $R\approx 1.37\times 10^{-3}$, 
to characterize the efficacy of the recoil expansion. 
Both SM and BSM couplings appear in 
${\Xi}$, $a_0$, and $b_{\mathrm{BSM}}$, namely~\cite{jackson57}
\begin{eqnarray}
{\Xi} &=& 1 + 3\lambda^2 + (g_S \epsilon_S)^2 + 3(4g_T \epsilon_T)^2\,, \\
\nonumber \\
a_0 &=& \frac{(1 - \lambda^2) - (g_S \epsilon_S)^2 + (4 g_T \epsilon_T)^2}
{(1 + 3\lambda^2) + (g_S \epsilon_S)^2 + 3(4 g_T \epsilon_T)^2}\,, \\
\nonumber \\
b_{\mathrm{BSM}} &=& \frac{2(g_S \epsilon_S) - 6\lambda(4 g_T \epsilon_T)}
{(1 + 3\lambda^2) + (g_S \epsilon_S)^2 + 3(4g_T \epsilon_T)^2}\,,
\label{eq:b_BSM}
\end{eqnarray}
where we employ Eq.~(\ref{eq:defST}). 

Our recoil-order expression for the term proportional to
$\epsilon/Rx \propto m_e/E_e$ appearing within the first set of square 
 brackets in the
differential distribution of  Eq.\ (\ref{eq:parent_a}) is equivalent to
the term labeled ``$b_\text{SM}$'' 
employed in Refs. \cite{Gudkov:2005bu,bhattacharya12}. 
 However,
it should be noted that the second-class currents $f_3$ and $g_2$
yield an additional $m_e/E_e$ term which is proportional to $(f_3 -
\lambda g_2)$.  Simply for the sake of notation, we label this term
``$b_\text{SCC}$'', where we then have, in summary,
\begin{eqnarray}
b_\text{SM} &=& -\frac{m_e}{M}
  \frac{1 + 2\lambda + \lambda^2 + 2\lambda(\kappa_p - \kappa_n)}
  {1 + 3\lambda^2}\,, \nonumber \\
b_\text{SCC} &=& 2\frac{m_e}{M}
  \frac{f_3 - \lambda g_2}{1 + 3\lambda^2}\,.
\end{eqnarray}

In polarized $\beta$ decay, the 
differential distribution relevant to a measurement of $A$ 
 is of the form
\begin{eqnarray}
\frac{d^3 \Gamma}{dE_e d\Omega_e} &\propto&
M^4 R^4 \beta x^2 (1-x)^2 
\frac{1}{(1 + \epsilon - 2Rx)^3} h(x) \nonumber \\
&& \times 
\left[1 + b_{\mathrm{BSM}} \frac{m_e}{E_e} + A\beta\cos\theta_e \right]\,,
\label{eq:parent_A}
\end{eqnarray}
where $\theta_e$ is the angle between the momentum of the electron and the 
polarization of the neutron. Here, too, the structure of this
expression follows that of Refs.~\cite{bilenski60} and \cite{jackson57}
in suitable limits and serves as a 
definition of $A$ in recoil order; note that we neglect recoil contributions
to $b_{\mathrm{BSM}}$, so that ``$b_{\mathrm{BSM}}$'' is the same
quantity here and in Eq.~(\ref{eq:parent_a}). 
The complete expression for $h(x)$ can be found in
Ref.~\cite{zhang01} and is in agreement with Ref.~\cite{bender68}.
To LO in the recoil expansion,  $h(x)$ is of the form
\begin{equation}
h(x) = g_V^2 + 3 g_A^2 + (g_S \epsilon_S)^2 +
3(4g_T \epsilon_T)^2 \,.
\end{equation}
Working to LO in the $S$ and $T$ terms and to NLO in the $V-A$ terms, 
\begin{widetext}
\noindent 
\begin{eqnarray}
\frac{h(x)}{(1 + \epsilon - 2 Rx)^3} &=& 
(1 + 3\lambda^2) 
\left[1 + 3 Rx  + Rx 
\left( \frac{4\lambda(1 + \kappa_p - \kappa_n)}{1 + 3\lambda^2} \right) -
2 R \left( \frac{\lambda^2 + \lambda + \lambda(\kappa_p - \kappa_n)}
{1 + 3\lambda^2}\right) 
\right. \nonumber \\
&& \left. - 4 R \left( \frac{\lambda g_2}{1 + 3\lambda^2} \right) 
- \frac{\epsilon}{Rx} \left( 
\frac{1 + 2\lambda + \lambda^2 + 2\lambda(\kappa_p - \kappa_n)} 
{1 + 3\lambda^2} \right) + 
2 \frac{\epsilon}{Rx} \left(\frac{f_3 - \lambda g_2}{1 + 3\lambda^2}\right)\right] \nonumber \\
&+& (\lambda^2 -1) \left(Rx - \frac{\epsilon^2}{Rx} \right) 
 +  (g_S \epsilon_S)^2 + 3(4g_T \epsilon_T)^2 
 \,, 
\label{spectrum}
\end{eqnarray}
and $A$ is of the form
\begin{eqnarray}
A &=& A_0 + \frac{1}{(1 + 3\lambda^2)^2}
\Bigg\{\frac{\epsilon}{Rx}\left[4\lambda^2(1-\lambda)(1 + \lambda +
  2f_2) + 4\lambda(1-\lambda)(\lambda g_2 - f_3)\right] \nonumber \\
&&~~~~~~~~~~~~~~~~~~~~~~~~ + R\left[\frac{2}{3}[1 + \lambda +
  2(f_2 + g_2)](3\lambda^2 + 2\lambda -1)\right]
  \nonumber \\ 
&&~~~~~~~~~~~~~~~~~~~~~~~~ + Rx\left[\frac{2}{3}(1 + \lambda + 2f_2)
  (1 - 5\lambda - 9\lambda^2 - 3\lambda^3) +
  \frac{4}{3}g_2 (1 + \lambda + 3\lambda^2 + 3\lambda^3)\right]
  \Bigg\},
\label{eq:asym_recoil}
\end{eqnarray}
with 
\begin{equation}
A_0 = \frac{2\lambda(1 - \lambda) + 2(4g_T \epsilon_T)^2 +
2(g_S \epsilon_S)(4g_T \epsilon_T)}{(1 + 3\lambda^2) + (g_S \epsilon_S)^2 +
3(4g_T\epsilon_T)^2}\,. 
\end{equation}
Our expressions in the context of $V-A$ theory 
agree with those of Ref.~\cite{holstein74} --- and with 
those of Ref.~\cite{bilenski60} if $f_3=g_2=0$. 
\end{widetext}

As a final topic we revisit the 
computation of the neutron lifetime and focus 
particularly on the role of recoil-order corrections. 
In the current state of the art \cite{czarnecki04,marciano06}, 
$V_{ud}$, ${\Xi}$, and $\tau$ are related by 
\begin{equation}
\tau = \frac{4908.7(1.9)~\text{s}}{|V_{ud}|^2 {\Xi}}\,,
\label{eq:tau_lambda_Vud}
\end{equation}
where ${\Xi} = 1 + 3\lambda^2$ in the absence of new physics. 
Employing $V_{ud} =0.97425$ and $\lambda=1.2701$ \cite{PDG} 
yields a lifetime of $885.6$ s. 
The numerical value reported in Eq.~(\ref{eq:tau_lambda_Vud}),
as per Ref.~\cite{marciano06}, 
incorporates an improved treatment of electroweak radiative effects, including
certain ${\cal O}(\alpha^2)$ contributions \cite{czarnecki04,marciano06}. 
Note that the calculation embeds a value of $g_A=1.27$ in matching
the short- and 
long-distance radiative corrections \cite{czarnecki04,marciano06}. 
The numerical value also includes the phase space factor $f$ which incorporates
the Fermi function and various recoil-order terms \cite{wilkinson82}; many
small terms involving hadronic couplings other than $g_V$ and $g_A$ can enter
in recoil order. Reference \cite{wilkinson82} analyzes neutron $\beta$ decay
to 0.001\% in precision and 
finds the latter corrections negligible save for the possibility of 
that from $g_2$. 
In that work \cite{wilkinson82} 
contributions proportional to $f_3$ are assumed to be strictly 
zero from CVC, though such can also be engendered by SM isospin violation. 
We can evaluate the various small contributions to the total decay rate
by integrating the terms of 
Eq.~(\ref{spectrum}) 
over the allowed phase space of Eq.~(\ref{eq:parent_A}). We denote
a contribution relative to that from ${\Xi}$ by $C_{g_i g_j}$, 
where $g_i$ and $g_j$ are the couplings it contains. 
Defining 
$W_0=E_e^{\rm max}/m_e$, $W=E_e/m_e$, $p_W=\sqrt{W^2-1}$, and finally 
\begin{equation}
I_m = \int_{1}^{W_0} dW p_W W (W_0 - W)^2 W^m \,,
\end{equation}
the small contributions to the decay rate, relative to that mediated
in LO by ${\Xi}$,  are 
\begin{eqnarray}
C_{g_Ag_2} &=& - \frac{\lambda g_2}{{\Xi}} \frac{m_e}{M} \left(
4 W_0 + 2 \frac{I_{-1}}{I_0} \right) \,,
\nonumber \\
C_{g_Vf_3} &=& \frac{2 f_3}{\Xi} \frac{m_e}{M} \frac{I_{-1}}{I_0} \,, \nonumber \\
C_{g_Af_2} &=& \frac{\lambda(\kappa_p-\kappa_n)}{\Xi} \frac{m_e}{M} 
\left(\frac{4I_1 - 2 W_0 I_0 -2 I_{-1}}{I_0} \right) \,, \nonumber \\
C_{g_V g_A} &=& \frac{\lambda}{\Xi} \frac{m_e}{M} 
\left(\frac{4I_1 - 2 W_0 I_0 -2 I_{-1}}{I_0}\right) \,, \nonumber \\
C_{g_A g_A} &=& - \frac{\lambda^2}{\Xi} \frac{m_e}{M} 
\left(\frac{2 W_0 I_0 + I_{-1}}{I_0} \right) \,,
\end{eqnarray}
where we note the appendix of Ref.~\cite{wilkinson82} for a useful tabulation
of the integrals $I_m$. 
In our current study, in which we assume that the entire range of
possible electron energies is experimentally accessible, 
both $C_{g_A f_2}$ and $C_{g_V g_A}$ 
{\it vanish} up to contributions 
nominally of ${\cal O}(\alpha R) \sim 1.0 \times 10^{-5}$ and
${\cal O}(R^2)\sim 1.9\times 10^{-6}$ 
in size, both of which are negligible at 0.001\% precision. 
Using the masses reported in Ref.~\cite{PDG}, 
specifically $M=939.565379\,{\rm MeV}$, 
$M^\prime=938.272046\,{\rm MeV}$, and $m_e=0.510998928\, {\rm MeV}$, 
the remaining terms evaluate to 
\begin{eqnarray}
C_{g_Ag_2} &=& -6.21 \times 10^{-3} \frac{\lambda g_2}{\Xi} 
\nonumber \\
C_{g_Vf_3} &=& 7.12\times 10^{-4} \frac{f_3}{\Xi}  \,, \nonumber \\
C_{g_A g_A} &=& -3.11\times 10^{-3} \frac{\lambda^2}{\Xi} \,. 
\end{eqnarray}
Using $\lambda=1.2701$ \cite{PDG} we find
$C_{g_A g_A}=-8.58 \times 10^{-4}$. 
If $\lambda$ changes within $\pm 0.10$, we note that 
$C_{g_A g_A}$ changes negligibly at the
precision to which we work,  so that we can regard 
$C_{g_A g_A}$ as a fixed constant in the optimizations
to follow. 
This particular contribution should already be embedded in 
the numerical constant of Eq.~(\ref{eq:tau_lambda_Vud}); however,
the terms involving second-class currents have not been. 
To include such small corrections in the lifetime we need
only replace $\Xi$ with $\Xi(1+C_{g_i g_j})$, so that 
to retain $g_2$, e.g., we write
\begin{equation}
\tau = \frac{4908.7~\mathrm{s}}{|V_{ud}|^2 
                          \left[\Xi  - (6.21\times 10^{-3})g_2\lambda\right]}
\label{eq:tau_g2}
\end{equation}

It is worth noting that finite experimental acceptance plays an important role 
in the assessment of the recoil corrections to the lifetime. 
For example, if the accessible electron kinetic 
energy were limited to the interval $[100, 700]$ keV from the allowed 
range of $[0,E_e^{\rm max} - m_e \approx 781.5]$ keV, 
then $C_{g_A f_2}$ and $C_{g_V g_A}$ would no longer vanish in 
${\cal O}(R)$.  
The integrals are still analytically soluble and 
evaluate to 
$C_{g_A f_2}=7.47\times 10^{-4}$ and $C_{g_V g_A}=2.02\times 10^{-4}$, 
where we use $f_2=(\kappa_p - \kappa_n)/2=1.8529450$ \cite{PDG} as well. 
Taken together, 
they yield a contribution some 100 times larger
than our earlier assessment, which was set by ${\cal O}(\alpha R)$. 
Including these effects in the
manner of Eq.~(\ref{eq:tau_g2}), they reduce the 
determined neutron lifetime for fixed $\lambda$ by $0.8$~s
to yield $884.8$ s. 
Such considerations 
differentiate neutron lifetime
experiments which (i) count surviving neutrons from those which (ii) 
count decay products. 
The existing tension between the latter, 
``in beam'' experiments and the former,  
``bottle'' experiments \cite{PDG,few11} 
--- though there is also tension between the results of 
the most precise 
bottle experiments \cite{PDG,few11} --- make the observation intriguing. 
However, the most precise in-beam neutron lifetime 
experiment \cite{Dewey:2003hc,Nico:2004ie} 
counts decay protons, rather than electrons, so that our
numerical analysis is not 
directly relevant. Indeed, in such experiments, there are no
threshold effects, and the entire proton recoil spectrum is
empirically accessible \cite{Dewey:2003hc,Nico:2004ie}. 
On the other hand, experimental concepts which detect
the decay electrons have been under development
\cite{huffman00,brome01}.

\section{Survey of Theoretical Uncertainties} 
\label{sec:uncertainties}

In the SM the corrections to the predictions of 
the CVC hypothesis, $g_V = 1$, $f_2 = (\kappa_p -
\kappa_n)/2$, and $f_3=0$ are parametrically known to be of 
${\cal O}(m_d - m_u)^2$ \cite{ademollogatto} for $g_V$ and of 
${\cal O}(m_d - m_u)$ 
for $f_2$ and $f_3$. The coupling $g_2$ does
not vanish under CVC, but rather from its $G$-parity properties; 
it, too, is nominally nonzero at ${\cal O}(m_d - m_u)$. 

The CVC hypothesis is assumed (i.e., $g_V = 1$)
in extracting values for $V_{ud}$ from measured $ft$ values in
superallowed $0^+ \rightarrow 0^+$ nuclear $\beta$ decay, and the
universality of $g_V$ in these decays has been tested to $1.3\times 10^{-4}$
at 68\% confidence level (CL) with a concomitant constraint of 
$m_e f_3/M g_V = -(0.0011\pm 0.0013)$ \cite{hardy09}, implying
$f_3$ is constrained to only $\mathcal{O}(1)$.  Direct
computation reveals the deviation of $g_V$ from unity to be smaller
still~\cite{Kaiser:2001yc}. Decay correlation measurements invariably
conflate tests of the CVC prediction for the weak magnetism form
factor \cite{gellmann58} with those which would limit second-class
currents; currently the CVC value of $f_2$ is tested to the level of
some 6\% \cite{severijns06,Sumikama:2011pr} at 68\% CL.  We note,
however, there has not, to date, been a published measurement of $f_2$
in neutron $\beta$ decay, such as, e.g., could be extracted from the linear 
energy dependence of $a$, or $A$ if $g_2=0$ is assumed.
The second-class couplings $f_3$ and $g_2$
have also not been probed experimentally in neutron $\beta$ decay.
The comparison of $ft$ values in mirror transitions can also test for
second-class currents; in that context the weak magnetism form factor
does not enter but an isospin-breaking additive correction from the
axial form factors can.  Such experiments give the strongest empirical
constraints on second-class currents
\cite{Minamisono:2001cd,Minamisono:2011zz} although additional
theoretical uncertainties enter.
A survey of nuclear $\beta$ decay data gives the limit 
$|g_2/f_2| < 0.1$ at 90\% CL, yielding 
$|g_2| < 0.2$  at 90\% CL \cite{Wilkinson:2000gx}. 

Some theoretical studies of $g_2$ exist, particularly in the case
of strangeness-changing 
transitions. A bag model estimate gives 
$g_2/g_V\sim 0.3$ \cite{donoghue82} in  $|\Delta S| =
1$ semileptonic transitions. 
More recently, non-zero second-class
currents have been observed in quenched lattice QCD calculations of form
factors which appear in the hyperon semileptonic decay 
$\Xi^0 \rightarrow \Sigma^+ {\ell} \bar \nu$, yielding $f_3/g_V = 0.14(9)$
and $g_2/g_A = 0.68(18)$ \cite{lattsu3}. Turning to the nucleon sector, 
we expect these estimates to be suppressed, crudely, by $m_d/m_s \sim 0.1$. 
This makes them nearly compatible in scale with 
the value for $g_2$ determined 
using QCD sum rule techniques, $g_2/g_A = -0.0152 \pm 0.0053$ \cite{shiomi96}.
  
The same lattice study has explored SU(3) breaking in the $f_2$
coupling as well, finding $[f_2/f_1]_{\Xi^0\to \Sigma^+}= 1.16(11)
\times [f_2/f_1]_{n\to p}$, a result somewhat different from the
predictions of common models of SU(3) breaking \cite{lattsu3}.
Applying a scaling factor of $m_d/m_s \sim 0.1$ we would suppose that
CVC breaking in $f_2$ is no larger than a few percent. It is notable
that the experimental limits on $f_2$, $f_3$, and $g_2$ are all rather
lax with respect to theoretical expectations of CVC breaking and SCC
from SM physics.

In what follows we explore the impact of a non-zero $g_2$, as well as of 
values of $f_3$ and 
$f_2$ which are not fixed precisely by the CVC prediction. 
These form factors all appear in recoil order; the usual first assumption
of uncorrelated errors suggests that the impact of these form factors
on the potential discovery of BSM physics ought be modest. This turns out to be 
not so because the fit parameters, rather, are highly correlated, as we shall see.

\section{Maximum Likelihood Analysis}
\label{sec:likelihood}

\subsection{Construction of the likelihood function}
\label{sec:likelihood_construction}

Having reviewed the formalism for neutron $\beta$ decay, the starting
point for our maximum likelihood analysis of neutron $\beta$ decay
observables is the 
frequentist $R$fit framework of the 
CKMfitter Group \cite{hocker01,charles05}; in future work we will
explore the other analysis schemes outlined by the CKMfitter Group
\cite{hocker01, charles05} and the UTfit Collaboration
\cite{ciuchini00,ciuchini01,Bona:2005vz,Bona:2006ah,Bona:2007vi}, 
such as Bayesian analyses.  We review
CKMfitter's $R$fit analysis in sufficient
detail to provide sufficient context for the discussion of our global
fit.  Complete details on the $R$fit statistical framework can, of
course, be found in their original papers
\cite{hocker01,charles05}. 

Our global fit includes two different types of
experimental observables: (i) results for angular correlation
coefficients as a function of (binned) electron energy and (ii) the
neutron lifetime.  We consider each of the bin-by-bin results for
the angular correlation coefficients to constitute a separate result.
Adopting the notation of the $R$fit framework, we label each of these
experimental measurements $x_{\text{exp},i}$.  Each of them 
are then compared with a corresponding
theoretical calculation of that quantity, $x_{\text{theo},i}$.  These
theoretical calculations are each a function of $N_\text{mod}$ model
parameters, the set of which we denote as $\{y_\text{mod}\}$.  Of these
$N_\text{mod}$ parameters, $N_\text{free} \leq N_\text{mod}$ are
 experimentally-accessible ``free parameters'' of the
model, the set of which we denote as $\{y_\text{free}\}$.  The
remaining $N_\text{calc} = N_\text{mod} - N_\text{free}$ ``calculated
parameters,'' for which there have been no prior experimental
measurements and which are not accessible in the current experiments (noting, e.g.,
second-class currents), must be calculated within the context of the 
model, subject to various assumptions.  The set of these calculated
parameters we denote as $\{y_\text{calc}\}$.

The set of experimental observables $\{x_\text{exp}\}$
includes binned-in-energy measurements of
$A_{\text{exp},i}(E_{e,j})$ and $a_{\text{exp},i}(E_{e,j})$ [where we
use the subscripts $i$ and $j$ to label the particular experiment and
the energy bin, respectively] and results for the neutron lifetime,
$\tau_{\text{exp},i}$, from different experiments,
\begin{eqnarray}
\left\{ x_\text{exp} \right\} &=& \left\{ A_{\text{exp,1}}(E_{e,1}),
  A_{\text{exp,1}}(E_{e,2}), \ldots ,
  \right. \nonumber \\
&&~~~~~\left. a_{\text{exp,1}}(E_{e,1}), a_{\text{exp,1}}(E_{e,2}), \ldots ,
  \right. \nonumber \\
&&~~~~~\left. \tau_{\text{exp,1}}, \tau_{\text{exp,2}}, \ldots \right\}.
\end{eqnarray}
These are then to be compared, one-by-one, with a corresponding set
of theoretical calculations,
\begin{eqnarray}
\left\{ x_\text{theo}(y_\text{mod}) \right\}
  &=& \left\{ A_{\text{theo,1}}(E_{e,1}),
  A_{\text{theo,1}}(E_{e,2}), \ldots ,
  \right. \nonumber \\
&&~~~~~\left. a_{\text{theo,1}}(E_{e,1}), a_{\text{theo,1}}(E_{e,2}), \ldots ,
  \right. \nonumber \\
&&~~~~~\left. \tau_{\text{theo,1}}, \tau_{\text{theo,2}}, \ldots \right\},
\end{eqnarray}
which depend on the set of $\{y_\text{mod}\}$ parameters.  Under the SM, 
the set of $\{y_\text{mod}\}$ parameters would include
\begin{equation}
\{y_\text{mod}\} = \{ \lambda, f_2, f_3, g_2, g_3, V_{ud} \}\,,
\end{equation}
though we note that $g_3$ does not appear in $\beta$ decay observables
computed through  NLO precision. 
Consequently we set $g_3=0$ in all that follows; we refer the reader
to the reviews of Refs.~\cite{Gorringe:2002xx,Bernard:2001rs} 
for information on this quantity. 
In the next section where we show results from example fits for
different scenarios, we define for each scenario which of the
$y_{\text{mod},i}$ parameters are to be considered a free or
calculated parameter. We note a subtlety in regards to $\lambda$: if
it is a fit parameter, rather than a calculated one, it can be modified
by the appearance of a right-handed coupling emergent from 
physics BSM~\cite{bhattacharya12}. 
Thus the determined value of $\lambda$ from a 
fit of neutron 
beta decay observables 
need not be equivalent to the calculated
value of $\lambda=g_A/g_V$. However, present lattice calculations of $g_A$
are of rather poorer precision than empirical determinations. 
The above sets could, of course, be trivially
expanded to accommodate measurements of other observables;
for example, the $\{x_\text{exp}\}$ set could include
measurements of the neutrino asymmetry $B$ and (direct) measurements of
the Fierz interference term $b_{\rm BSM}$ (via spectral shape measurements),
and/or the $\{y_\text{mod}\}$ set could include $b_{\rm BSM}$.

As per the usual prescription \cite{PDG}, we define our $\chi^2$ function
in terms of a likelihood function $\mathcal{L}(y_\text{mod})$ for the
$\{y_\text{mod}\}$ parameter set as 
\begin{equation}
\chi^2(y_\text{mod}) = -2\ln \mathcal{L}(y_\text{mod})\,.
\label{eq:chi2}
\end{equation}
Following the $R$fit framework \cite{hocker01, charles05}, we
define $\mathcal{L}(y_\text{mod})$ to be the product
of an ``experimental likelihood'' function, $\mathcal{L}_\text{exp}$,
and a ``theoretical likelihood'' function, $\mathcal{L}_\text{calc}$,
\begin{eqnarray}
\mathcal{L}(y_\text{mod}) &=& 
\mathcal{L}_\text{exp}(\{x_\text{exp}\}, \{x_\text{theo}(y_\text{mod})\}) 
\nonumber \\
&& \times \mathcal{L}_\text{calc}(\{y_\text{calc}\}).
\end{eqnarray}

The experimental likelihood function is defined in the usual way to be
the product of the likelihood functions for each of the $N_\text{exp}$
experimental results,
\begin{equation}
\mathcal{L}_\text{exp}(\{x_\text{exp}\}, \{x_\text{theo}\}) =
\prod_{i=1}^{N_\text{exp}} \mathcal{L}_{\text{exp},i} \,,
\end{equation}
where, in the ideal case, the individual likelihood functions
are taken to be Gaussian,
\begin{equation}
\mathcal{L}_{\text{exp},i} = \frac{1}{\sqrt{2\pi\sigma_{\text{exp},i}^2}}
\exp\left[-\frac{(x_{\text{exp},i} - x_{\text{theo},i})^2}
{2\sigma_{\text{exp},i}^2}\right]\,,
\end{equation}
where $\sigma_{\text{exp},i}$ denotes the statistical uncertainty of
the $i$th experimental result.  Of course, the individual
experimental likelihood functions must account for systematic errors,
and the formalism for the inclusion of such within the context of the
$R$fit framework is described in detail in Refs.~\cite{hocker01,charles05}.
However, a detailed discussion of the impact of experimental
systematic errors is beyond the scope of this paper, as the focus of
our first paper is on the  statistical impact of a global fit and
the limitations on such from theoretical uncertainties.

In principle, a theoretical likelihood function could similarly be
defined as the product of likelihood functions for each of the
$y_{\text{calc},i}$ calculated parameters,
\begin{equation}
\mathcal{L}_\text{calc}(\{y_\text{calc}\}) =
\prod_{i=1}^{N_\text{calc}} \mathcal{L}_{\text{calc},i}\,,
\end{equation}
and under the assumption that the $y_{\text{calc},i}$ values are
Gaussian distributed, the theoretical likelihood would, by definition,
contribute to the $\chi^2$.  Such a formulation might not be 
appropriate for the treatment of the $y_{\text{calc},i}$ parameters,
for which the underlying probability distributions are certainly not
known.  However, 
it may well be possible 
to bound the value of each $y_{\text{calc},i}$ parameter
on theoretical grounds, such that the
parameter may reasonably assume any value over an allowed range of
$[y_{\text{calc},i} - \delta y_{\text{calc},i}, y_{\text{calc},i} +
\delta y_{\text{calc},i}]$. It would be highly unlikely that the true value
of the parameter would fall outside of this range.

Thus, given the lack of knowledge on the underlying distributions of
the $y_{\text{calc},i}$ parameters, the proposal of the $R$fit
scheme \cite{hocker01,charles05} is to redefine the $\chi^2$ function
so that the theoretical likelihood does not contribute to
the $\chi^2$, while the $y_{\text{calc},i}$ parameters are permitted
to vary freely within their pre-defined allowed ranges.  In particular,
the $\chi^2$ is re-defined to be
\begin{equation}
\chi^2 = \sum_{i=1}^{N_\text{exp}}\left(\frac{x_{\text{exp},i} -
x_{\text{theo,i}}}{\sigma_{\text{exp},i}}\right)^2 -
2\ln\mathcal{L}_\text{calc}(\{y_\text{calc}\}),
\label{eq:chi2_modified}
\end{equation}
where
\begin{equation}
\! -2\ln\mathcal{L}_\text{calc}(\{y_\text{calc}\}) \equiv \!
  \left\{\begin{array}{ll}
0,& \forall\, y_{\text{calc},i} \in
  [y_{\text{calc},i} \pm \delta y_{\text{calc},i}] \\
\infty,& \text{otherwise}
\end{array}
\right. \,.\nonumber
\end{equation}
Thus, under the $R$fit scheme, each of the $y_{\text{calc},i}$
parameters are bounded, but all possible values of the parameters
within their pre-defined ranges are treated equally.  That is, the
value of $\chi^2$ is scanned over the available $\{y_\text{free}\}$
parameter space, while the values of the $\{y_\text{calc}\}$ parameters are 
permitted to vary freely over their pre-defined ranges \textit{at each point}
in the $\{y_\text{free}\}$ parameter space.  Thus, the central 
challenge of such an analysis in Refs.~\cite{hocker01,charles05} is to
define the $[y_{\text{calc},i} \pm \delta y_{\text{calc},i}]$ allowed
ranges carefully because: (i) the fit results for the 
$\{y_\text{free}\}$ parameters can be interpreted as valid only if the
``true'' values for the $\{y_\text{calc}\}$ parameters fall within the
allowed ranges; and (ii) choosing the allowed ranges to be too wide
(i.e., too conservative) could mask the discovery of new physics.

After construction of the $\chi^2$, a global fit can be then be
pursued under two different types of analyses: (i) determining values
for the SM parameters; and (ii) assessing the validity of
the SM.

\subsection{Determining standard model parameters}
\label{sec:likelihood_minimization}

Here the goal is neither to assess the validity of the SM nor
to search for evidence of new physics.  Instead, the SM is
assumed to be valid, and the global fit is employed, optimally, to
determine values for all of the $\{y_\text{mod}\}$ parameters.  In
this case, the minimum value of $\chi^2(y_\text{mod})$, computed
according to Eq.\ (\ref{eq:chi2}), is obtained by allowing all of the
$N_\text{mod}$ parameters to freely vary.  The resulting minimum value
is denoted $\chi^2(y_\text{mod})_\text{min}$.  Confidence
levels, $\mathcal{P}(y_\text{mod})$, on the values of the parameters
obtained at $\chi^2(y_\text{mod})_\text{min}$ are calculated according
to
\begin{equation}
\mathcal{P}(y_\text{mod}) = \text{Prob}(\Delta\chi^2(y_\text{mod}),
N_\text{dof}),
\end{equation}
where, as usual,
\begin{equation}
\Delta\chi^2(y_\text{mod}) = \chi^2(y_\text{mod}) -
\chi^2(y_\text{mod})_\text{min},
\end{equation}
and Prob(\ldots) denotes the probability for a value of $\chi^2 >
\Delta\chi^2(y_\text{mod})$ for $N_\text{dof}$ degrees of freedom.

However, in practice, it may not be possible, or feasible, to
determine values for all of the $\{y_\text{mod}\}$ parameters. 
The $\{y_\text{calc}\}$ parameters, notably the second-class
couplings, $f_3$ and $g_2$, and the induced pseudoscalar coupling 
$g_3$, serve as examples. 
 We let $\{y_a\}$ denote the $N_a \leq N_\text{mod}$ subset of
$\{y_\text{mod}\}$ parameters for which the goal is to determine
confidence levels; the remainder of the $N_\mu = N_\text{mod} - N_a$
parameters are denoted $\{y_\mu\}$.  Note that the set of $\{y_\mu\}$
parameters may not be identical to the set of $\{y_\text{calc}\}$
parameters.

Confidence levels on the $\{y_a\}$ parameter set are then determined
by first computing \textit{at each point} in the $\{y_a\}$ parameter space the
minimal value of the $\chi^2$ function,
$\chi^2(\{y_a\};\{y_\mu\})_\text{min}$, obtained by allowing the
$\{y_\mu\}$ parameters to vary.  The minimal value of $\Delta\chi^2$
at that point in the $\{y_a\}$ parameter space is then computed
according to
\begin{equation}
\Delta\chi^2(\{y_a\}) = \chi^2(\{y_a\};\{y_\mu\})_\text{min} -
\chi^2(y_\text{mod})_\text{min}.
\end{equation}
The confidence levels are then obtained from
\begin{equation}
\mathcal{P}(\{y_a\}) = \text{Prob}(\Delta\chi^2(\{y_a\}),N_\text{dof}).
\end{equation}

\subsection{Assessing the standard model}
\label{sec:likelihood_assessing}

Under the minimization scheme just described for the determination of
SM parameters, the SM is assumed 
to be valid by definition.  In principle, a test statistic for assessing the
validity of the SM would be the value of
$\chi^2(y_\text{mod})_\text{min}$ obtained when all of the
$N_\text{mod}$ are varied, where a confidence level on the SM 
could be defined as
\begin{equation}
\mathcal{P}(\text{SM}) \leq \text{Prob}(\chi^2(y_\text{mod})_\text{min},
N_\text{dof})\,.
\end{equation}

\begin{table*}
\caption{Summary of parameters for the ``Standard Model'' and
``New Physics'' Monte Carlo pseudodata sets. We note under CVC
that $f_2=(\kappa_p-\kappa_n)/2=1.8529450$ \cite{PDG}.}
\begin{ruledtabular}
\begin{tabular}{lccccccc}
Input Parameters& $\lambda$& $f_2$& $f_3$& $g_2$& $g_3$& $g_S\epsilon_S$&
  $g_T\epsilon_T$ \\ \hline
Standard Model& PDG: 1.2701& CVC: $(\kappa_p-\kappa_n)/2$& 0& 0& 0&
  0& 0 \\
New Physics&    PDG: 1.2701& CVC: $(\kappa_p-\kappa_n)/2$& 0& 0& 0&
  0& $1.0 \times 10^{-3}$ \\ \hline
Calculated Parameters& $\Xi$& $a_0$& $A_0$& $b_\text{BSM}$& $\tau$ \\ \hline
Standard Model& 5.83946& $-0.105002$& $-0.117495$& 0&          885.631 s& \\
New Physics&    5.83951& $-0.104998$& $-0.117489$& $-0.00522$& 885.624 s& \\
\end{tabular}
\end{ruledtabular}
\label{tab:simulation_input_parameters}
\end{table*}

In practice, an assessment of the validity of the SM can
be obtained via a Monte Carlo calculation according to the following
scheme.  Values for the set of $\{x_\text{exp}\}$ experimental results
are sampled in the Monte Carlo from their corresponding set of theoretical
expressions, $\{x_\text{theo}\}$, assuming the fitted values for the
$\{y_\text{mod}\}$ parameter set.  For each set of $\{x_\text{exp}\}$
values, a $\chi^2$ value is computed, as before, by allowing the
$y_\text{mod}$ parameters to vary.  This is repeated, and from this a
(normalized) distribution of $\chi^2$ values, $p(\chi^2)$, is
constructed.  A confidence level for the SM is then
deduced from this distribution according to
\begin{equation}
\mathcal{P}(\text{SM}) \leq {\int\limits_{\chi^2 \geq
\chi^2(y_\text{mod})_\text{min}}} \!\!\!\!\!\!\!\!\!\! p(\chi^2)~d\chi^2.
\end{equation}

\section{Example Fit Scenarios}
\label{sec:example_fits}

We now illustrate, via several examples of $n$Fitter fits to
Monte Carlo pseudodata, the impact that simultaneous fits to the
energy dependence of the $a$ and $A$ angular correlation coefficients
have on an assessment of the validity of the SM and
the extent to which theoretical uncertainties can limit such an
assessment.

\subsection{Monte Carlo data sets and statistics}
\label{sec:example_fits_monte_carlo_data_sets}

We generated Monte Carlo pseudodata 
by sampling the relevant
recoil-order differential distributions for measurements of $a$ and
$A$, Eqs.\ (\ref{eq:parent_a}) and (\ref{eq:parent_A}), respectively, 
employing the complete expression for $h(x)$ in Ref.~\cite{zhang01}
in the latter case --- only the terms of Eq.~(\ref{spectrum}) 
are appreciably nonzero. 
We generated two different data sets.  The first data set, which we
term our ``Standard Model'' data set, consists of $5 \times 10^9$
simulated events for a measurement of $a$, and a separate data set
consisting of $5 \times 10^9$ simulated events for a measurement of
$A$.  Both data sets employ the current Particle Data Group average
value for $\lambda = 1.2701$ \cite{PDG} and the CVC value for 
$f_2=(\kappa_p-\kappa_n)/2=1.8529450$ \cite{PDG}, 
 and assume all of the 
other small terms are zero, $f_3 = g_2 = g_3 = 0$.

The second data set, which we term our ``New Physics'' data set, again
consists of $5 \times 10^9$ simulated events for a measurement of
$a$, and $5 \times 10^9$ events for a measurement of $A$.  
This data set is identical to the earlier one, save for the
inclusion of a nonzero value for a tensor coupling, namely, 
$g_T \epsilon_T = 1.0 \times 10^{-3}$, 
close to the strongest empirical limit on this quantity, which comes 
from a Dalitz study of 
radiative pion decay \cite{Bychkov:2008ws,Cirigliano:2013xha}.  
Specifically, we note the extracted 90\% CL limit on 
${\rm Re}(\epsilon_T)$ \cite{Bychkov:2008ws,Cirigliano:2013xha} 
can be combined 
with $g_T=1.05 (35)$ \cite{bhattacharya12}, 
or $g_T< 1.4$, 
both in the 
${\overline {\rm MS}}$ scheme at a renormalization scale of 2 GeV, 
to yield 
$-1.5 \times 10^{-3} < g_T {\rm Re}(\epsilon_T) < 1.9 \times 10^{-3}$. 
The most stringent limit on $g_S {\rm Re}(\epsilon_S)$, which 
comes from the 
analysis of $0^+\to 0^+$ nuclear decays \cite{hardy09}, is of 
a comparable magnitude. 
However, as can be seen from
Eq.\ (\ref{eq:b_BSM}), for approximately equal values of
$g_S\epsilon_S$ and $g_T\epsilon_T$, $b_\text{BSM}$ is significantly
more sensitive to tensor couplings; hence, we have decided to illustrate
our methods using a non-zero tensor coupling exclusively.  A
summary of our input parameters and the resulting values for $\Xi$,
$a_0$, $A_0$, $b_\text{BSM}$, and $\tau$ 
are given in Table 
\ref{tab:simulation_input_parameters}.
Note that we calculate $\tau$ as per 
Eq.\ (\ref{eq:tau_lambda_Vud}), assuming the central value
for the superallowed $0^+ \rightarrow 0^+$ value of $V_{ud} =
0.97425(22)$ \cite{PDG}. 

Our pseudodata consists of $5 \times 10^9$ events, because such 
is needed for the anticipated level of statistical precision in the
most ambitious of the next generation of decay correlation 
experiments. Specifically, 
we would be able to determine the
value of $\lambda$ in an $a$ measurement to 0.010\% (equivalent to a
0.037\% determination of $a_0$) and to 0.008\% in an $A$ measurement
(0.032\% determination of $A_0$). 
The stated goals on $a$ and $A$ in the upcoming PERC experiment
\cite{Dubbers:2007st} are to achieve 
statistical and systematic errors on the level of 0.03\%.

The numerical results presented hereafter employ the full $E_e$
energy range (i.e., kinetic energies $0 \leq T_e \leq T_0$).  Of
course, in a real experiment, the lower energy range will necessarily
be in excess of zero 
due to hardware thresholds and/or analysis cuts.  Also,
considerations of systematics may limit the upper energy range
because energy loss effects become disproportionately more important
as the electron energy increases, see,
e.g., Refs.~\cite{liu10,plaster12,mendenhall13}.  Such details would, of
course, be included in a global fit to actual data; the point of this
paper is to illustrate the method. 

\subsection{Examples: Fits to the Standard Model data set}
\label{sec:example_fits_example_standard_model}

\begin{figure*}
\includegraphics[scale=0.87]{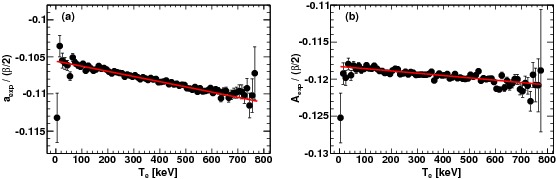}
\caption{(Color online) Simulated data from the Standard Model data
  set for $a_\text{exp}/(\frac{1}{2}\beta)$ [panel (a)] and
  $A_\text{exp}/(\frac{1}{2}\beta)$ [panel (b)] plotted as a function
  of $T_e$.  The solid red line is the result of a simultaneous fit
to the $a$ and $A$ data.}
\label{fig:standard_model_lambda}
\end{figure*}

For our first example, we consider fits to the Standard Model data
set.  As an illustration of our methods, we  show 
our simulated data for $a$ and
$A$ as a function of $T_e$ in Fig.\ \ref{fig:standard_model_lambda}, 
employing 79 10-keV bins from 0--790 keV, as the
endpoint is $T_0 = 781.5$ keV. 
We plot the
``experimental asymmetries'' $a_\text{exp}$ and $A_\text{exp}$, scaled
by the nominal $\frac{1}{2}\beta$ electron-energy dependence of the
asymmetries, where the factor of $\frac{1}{2}$ results from the
angular integral $\langle \cos\theta_{e,e\nu} \rangle = \frac{1}{2}$ on 
a hemisphere. 
These experimental asymmetries $a_\text{exp}$ and $A_\text{exp}$
are calculated from the simulated data in a manner similar to how
actual experimental data would be analyzed in a typical
``forward/backward'' asymmetry measurement (see, e.g.,
\cite{plaster08}), where
\begin{eqnarray}
a_\text{exp} &\equiv& \frac{N(\cos\theta_{e\nu}>0) -
  N(\cos\theta_{e\nu}<0)} {N(\cos\theta_{e\nu}>0) +
  N(\cos\theta_{e\nu}<0)} \nonumber \\ &=&
\frac{1}{2}\beta\frac{a_1}{1 + b_\text{BSM}\frac{m_e}{E_e} +
  \frac{1}{3}a_2\beta^2}, \label{eq:a_exp} \\
A_\text{exp} &\equiv& \frac{N(\cos\theta_e>0)
  - N(\cos\theta_e<0)} {N(\cos\theta_e>0) + N(\cos\theta_e<0)}
\nonumber \\ &=& \frac{1}{2}\beta\frac{A}{1 + b_\text{BSM}\frac{m_e}{E_e}}.
\label{eq:A_exp}
\end{eqnarray}
Sensitivity to $b_{\rm BSM}$ from $A(E)$ and $a(E)$ have
been previously considered 
by Refs.~\cite{gluck95,bhattacharya12, hickerson13}. 
In a real experiment the effects of ${\cal O}(\alpha)$ radiative 
corrections \cite{Shann:1971fz} would have to be removed to interpret
$A_\text{exp}$ in terms of the simple theoretical expressions we employ,
noting Eqs.~(\ref{eq:arecoil}) and (\ref{eq:asym_recoil}), in our fits. 
We avoid this now for simplicity, and we are able to 
do so because said correction incurs no additional hadronic uncertainty. 
Moreover, for similar reasons 
we drop the $a_2$ term from our fits as well; they are 
simply trivially small. 
The fits shown in Fig.\ \ref{fig:standard_model_lambda} are the result
of a simultaneous fit to the $a$ and $A$ data, in which
$\{x_\text{exp}\} = \{a,A\}$, noting $\{a,A\}$ is shorthand
for the complete set of the binned-in-energy results for
$a_\text{exp}$ and $A_\text{exp}$, and $\{y_a\} = \{\lambda\}$. 
We fix $f_2$ to its CVC value and set 
all second-class couplings to zero, so that $b_\text{BSM}$ vanishes. 
As a validation of
  our methods, the fit result for $\lambda = 1.27009(8)$ agrees with
  the input value to within $-0.1\sigma$ with a 
  $\chi^2_\text{min}/N_\text{dof} = 135.3/157$, yielding a perfectly
  acceptable $\text{Prob}(\chi^2 > \chi^2_\text{min}) = 0.89$.

Relaxing the assumption that second class currents are zero, we 
apply the $R$fit scheme to a fit in which $\{y_a\} =
\{\lambda\}$, and $f_3$ and $g_2$ comprise the
$\{y_\mu\}$ parameter set, which are then permitted to 
  vary simultaneously over some prescribed range, 
as per the prescription 
discussed in Sec.\ \ref{sec:likelihood_minimization}.  Of the other
  $\{y_\text{mod}\}$ parameters, $f_2$ is again fixed to its CVC
  value, and $b_\text{BSM}$ is fixed to zero.  The
  resulting 68.3\% CL on $\lambda$ for different
  assumptions on the permitted theory ranges for $f_3$ and/or $g_2$
  are compared in Table \ref{tab:standard_model_lambda_f3_g2}.  
  Note that we determine a 68.3\% CL as per the requirement 
    $\Delta\chi^2(\{y_a\}) = \chi^2(\{y_a\};\{y_\mu\})_\text{min} -
    \chi^2(y_\text{mod})_\text{min} = 1$, where in this case $\{y_a\} =
    \{\lambda\}$ and $\{y_\mu\} = \{f_3,g_2\}$. 
Referring to Table \ref{tab:standard_model_lambda_f3_g2}, 
unless $g_2$ can be constrained to 
  $\mathcal{O}(0.1)$, theory uncertainties in $g_2$ would limit the
  precision to which $\lambda$ can be extracted from experiments
  aiming to measure $a$ and $A$ to the level of 0.03\%.  Even at this
  level, the range of the 68.3\% CL on $\lambda$ is
  $\sim 50$\% larger than the case in which second class currents are
  taken to be exactly zero.  
In contrast the fits are almost
  completely insensitive to the value of $f_3$; this is because 
the latter appears 
only in the $\epsilon/Rx$ terms. 

\begin{table}
\caption{Fitted values for $\lambda$ from a simultaneous fit to the
$\{a,A\}$ Standard Model data set under the inclusion of the
  indicated theoretical uncertainties in $f_3$ and $g_2$.  The fitted
  values for $\lambda$ are defined by the location of the overall
  $\chi^2_\text{min}$, while the 68.3\% CL is defined by
  $\Delta\chi^2 = 1$.  An asymmetric CL about the
  $\chi^2_\text{min}$ value is indicated by asymmetric upper $(+)$ and
  lower $(-)$ error bars.  Recall that the input value for $\lambda =
  1.2701$ \cite{PDG}.}
\begin{ruledtabular}
\begin{tabular}{ccl}
$f_3$ range& $g_2$ range& Fitted $\lambda$ \\ \hline
$0$& $0$&                   $1.27009(8)$ \\ \hline
$0$& $[-0.1,0.1]$&          $1.27036(+7)(-17)$ \\
$[-0.1,0.1]$& $0$&          $1.27009(8)$ \\ \hline
$0$& $[-0.5,0.5]$&          $1.27056(31)$ \\
$[-0.5,0.5]$& $0$&          $1.27007(+11)(-7)$ \\ \hline
$[-0.1,0.1]$& $[-0.1,0.1]$& $1.27035(+9)(-17)$ \\
$[-0.5,0.5]$& $[-0.5,0.5]$& $1.27054(+32)(-31)$ \\
\end{tabular}
\end{ruledtabular}
\label{tab:standard_model_lambda_f3_g2}
\end{table}

\begin{table}
\caption{Fitted values for $\lambda$ and $g_2$ from two-parameter
simultaneous, $\{a,A\}$, and individual, $\{a\}$ or $\{A\}$,
fits to the Standard Model data set.  The fitted values for
$\lambda$ and $g_2$ are defined by the location of the overall
$\chi^2_\text{min}$, while the 68.3\% CL is defined by
$\Delta\chi^2 = 2.30$.  We use the inputs $\lambda = 1.2701$ \cite{PDG}
and $g_2 = 0$.  The column ``Prob'' indicates the  Prob$(\chi^2 >
\chi^2_\text{min})$.}
\begin{ruledtabular}
\begin{tabular}{lccccc}
Fit& Fitted $\lambda$& Pull& $g_2$& $\chi^2_\text{min}/N_\text{dof}$&
  Prob \\ \hline
$\{a,A\}$& 1.27056(48)& $0.97\sigma$& 0.174(171)& 133.0/156& 0.91 \\
$\{a\}$& 1.294(40)& $0.60\sigma$& 6.7(11.3)& 54.4/77& 0.98 \\
$\{A\}$& 1.2708(14)& $0.50\sigma$& 0.28(65)& 77.8/77& 0.45 \\ \hline
Fit& \multicolumn{4}{c}{Correlation coefficient $\rho_{\lambda g_2}$} \\ \hline
$\{a,A\}$& \multicolumn{4}{c}{0.97} \\
$\{a\}$&   \multicolumn{4}{c}{1.00} \\
$\{A\}$&   \multicolumn{4}{c}{0.99} \\
\end{tabular}
\end{ruledtabular}
\label{tab:standard_model_lambda_g2}
\end{table}

Alternatively, in the absence of a theory bound on $g_2$, one could
fit directly for $\lambda$ and $g_2$. In what follows we neglect the
small $f_3$ contribution, setting $f_3=0$, which is reasonable, as we show in 
Table \ref{tab:standard_model_lambda_f3_g2}.  Results from a
fit in which $\{y_a\} = \{\lambda,g_2\}$, with no $y_\mu$
parameters, 
are
shown in Table \ref{tab:standard_model_lambda_g2}, 
where the errors on
$\lambda$ and $g_2$ are defined by $\Delta\chi^2 = 2.30$, i.e., the
68.3\% CL for a joint fit of two free parameters \cite{PDG}.  As one 
would expect, the error on $\lambda$ from a two-parameter
simultaneous fit to the $\{a,A\}$ data set is a factor of $\sim 6$
larger than that from a single-parameter fit for $\lambda$ alone; and,
it is worth noting that $g_2$ can be determined from such a fit to
$\mathcal{O}(0.2)$.  However, what is more interesting are the errors
on $\lambda$ and $g_2$ extracted from a fit to the $\{a\}$ data set
alone: the error on $\lambda$ is a factor of $\sim 300$ larger than
that from a single-parameter fit for $\lambda$ alone, and the error on
$g_2$ is of $\mathcal{O}(10)$.  The origin of this effect 
is clear: $g_2$ appears in $a_1$ only in the
expansion for $a$, Eq.\ (\ref{eq:arecoil}), namely via the combination
$2(\lambda - \lambda^3)(1 + g_2)$.  Therefore, $\lambda$ and $g_2$ are
directly correlated in a small recoil-order term in $a$, with only the
fitted value of $a_0$ ultimately limiting the $\Delta\chi^2$ range
and, hence, the $\lambda$ and $g_2$ errors.  In Table
\ref{tab:standard_model_lambda_g2}, we also show the ``Pull,'' which
we define as ${\rm Pull} = (x_{\rm fit} - x_{\rm input})/(\sigma_{\rm
fit})$ for a parameter $x$. In the event of asymmetric errors, we
average the two errors
to form $\sigma_{\rm fit}$. 
For completeness, the correlation
coefficients $\rho_{\lambda g_2}$ from the fits are also given. 
Thus, this example succinctly illustrates the 
necessity of a stringent theoretical bound on the value of $g_2$ for an
interpretation of $a$ measurements in the context of an assessment of
the $V-A$ structure of the SM.

As a final example of fits to our Standard Model data set, we
consider the implications of CVC breaking on the value of the weak
magnetic form factor $f_2$, by defining $\{y_a\} = \{\lambda\}$
and $\{y_\mu\} = \{f_2\}$ under different assumptions on the
permitted theory range for $f_2$.  The results are summarized
in Table \ref{tab:standard_model_f2}.  As can be seen there,
the impact of a $\pm 2$\% breaking on $f_2$ is comparable
to an $\mathcal{O}(0.1)$ uncertainty in $g_2$.

\begin{table}
\caption{Fitted values for $\lambda$ from a simultaneous fit to the
  $\{a,A\}$ Standard Model data set under different assumptions on the
  range of CVC breaking for the value of $f_2$.  The fitted values for
  $\lambda$ are defined by the location of the overall
  $\chi^2_\text{min}$, while the 68.3\% CL is defined by
  $\Delta\chi^2 = 1$.  We use the input $\lambda =
  1.2701$ \cite{PDG}.}
\begin{ruledtabular}
\begin{tabular}{cl}
$f_2$ range & Fitted $\lambda$ \\ \hline
CVC exact& 1.27009(8) \\
$\pm 1$\%& 1.27009(9) \\
$\pm 2$\%& 1.27011(11) \\
$\pm 5$\%& 1.27016(15) \\
\end{tabular}
\end{ruledtabular}
\label{tab:standard_model_f2}
\end{table}

\subsection{Examples: Fits to the New Physics data set}
\label{sec:example_fits_example_new_physics}

As our second example, we consider fits to the New Physics data
set.  As can be seen in Table \ref{tab:simulation_input_parameters},
the values of $a_0$ and $A_0$ in the New Physics data set differ from
their values in the Standard Model data set by only 0.004\% and
0.005\%, respectively. Therefore, the impact of any new physics 
from scalar and tensor interactions on the measured values
of $a$ and $A$ will be via a ``dilution'' to the experimental
asymmetries $a_\text{exp}$ and $A_\text{exp}$ from a non-zero Fierz
term $b_\text{BSM}$ appearing in the denominators of
Eqs.\ (\ref{eq:a_exp}) and (\ref{eq:A_exp}), respectively.
Accordingly, we now expand our $\{y_\text{mod}\}$ parameter
set to include $b_\text{BSM}$.

Results from a single parameter fit in which $\{y_a\} = \{\lambda\}$
only and an empty $\{y_\mu\}$ parameter set are shown in Table
\ref{tab:new_physics_lambda}.  As can be seen, these fits still
yield excellent values for $\chi^2_\text{min}$; however, not
surprisingly, there are significant pulls on the fitted values for
$\lambda$ from the input value.  The origin of these pulls can be
easily understood (and, indeed, was first noted by Ref.\ 
\cite{bhattacharya12}) by inspecting the functional forms for
$a_\text{exp}$ and $A_\text{exp}$ in terms of $a$ and $A$.  For small
$b_\text{BSM}$, $\alpha_\text{exp} \approx \alpha(1 -
b_\text{BSM}m_e/E_e)$ with $\alpha = a$ or $A$, whereas the slope
of the $\epsilon/Rx \propto m_e/E_e$ dependence of $a_1$ and $A$ is negative
in $\lambda$ and nearly linear; the $m_e/E_e$
contributions to $a_\text{exp}$ and $A_\text{exp}$ from $b_\text{BSM}$
and $\lambda$ are of the same sign and slope.  Therefore, the presence
of a non-zero $b_\text{BSM}$ would not result in poor $\chi^2$ values
in fits to the $a$ and $A$ energy dependence.

\begin{table}
\caption{Fitted values for $\lambda$ from simultaneous, $\{a,A\}$, and
  individual, $\{a\}$ or $\{A\}$, fits to the New Physics
  data set, in which $\lambda$ was the only free parameter.  The
  fitted values for $\lambda$ are defined by the location of the
  overall $\chi^2_\text{min}$ in each case, 
  while the 68.3\% CL is defined by
  $\Delta\chi^2 = 1$.  We use the input $\lambda =
  1.2701$ \cite{PDG}.  The column ``Prob'' indicates the Prob$(\chi^2 >
  \chi^2_\text{min})$.}
\begin{ruledtabular}
\begin{tabular}{lcccc}
Fit& Fitted $\lambda$& Pull& $\chi^2_\text{min}/N_\text{dof}$& Prob \\ \hline
$\{a,A\}$& 1.27115(8)& $13.1\sigma$& 153.6/157& 0.56 \\
$\{a\}$& 1.27135(13)& $9.6\sigma$& 75.0/78& 0.57 \\
$\{A\}$& 1.27103(10)& $9.2\sigma$& 74.7/78& 0.59
\end{tabular}
\end{ruledtabular}
\label{tab:new_physics_lambda}
\end{table}

\begin{table}
\caption{Fitted values for $\lambda$ from simultaneous, $\{a,A\}$, and
  individual, $\{a\}$ or $\{A\}$, fits to the New Physics data set,
  under the inclusion of the indicated theoretical uncertainties in
  $f_3$ and $g_2$.  The fitted values for $\lambda$ are defined by the
  location of 
 $\chi^2_\text{min}$, while the 68.3\%
  CL is defined by $\Delta\chi^2 = 1$.  An asymmetric CL about
  the $\chi^2_\text{min}$ value is indicated by asymmetric upper $(+)$
  and lower $(-)$ error bars.  We use the input 
  $\lambda = 1.2701$ \cite{PDG}. 
 The column ``Prob'' indicates the Prob$(\chi^2 >
  \chi^2_\text{min})$.}
\begin{ruledtabular}
\begin{tabular}{lcccc}
\multicolumn{5}{c}{$f_3 \in [-0.1,0.1]$ and $g_2 \in [-0.1,0.1]$} \\
Fit& Fitted $\lambda$& Pull& $\chi^2_\text{min}/N_\text{dof}$& Prob \\ \hline
$\{a,A\}$& $1.27088(+19)(-7)$& $6.0\sigma$& 151.9/157& 0.60 \\
$\{a\}$& $1.27170(+13)(-83)$& $3.3\sigma$& 75.0/78& 0.58 \\
$\{A\}$& $1.27124(+10)(-27)$& $6.2\sigma$& 73.9/78& 0.61 \\ \hline
\multicolumn{5}{c}{$f_3 \in [-0.5,0.5]$ and $g_2 \in [-0.5,0.5]$} \\
Fit& Fitted $\lambda$& Pull& $\chi^2_\text{min}/N_\text{dof}$& Prob \\ \hline
$\{a,A\}$& $1.27072(+30)(-32)$& $2.0\sigma$& 151.4/157& 0.61 \\
$\{a\}$& $1.27311(+15)(-365)$& $\sim 0$& 74.8/78& 0.58 \\
$\{A\}$& $1.27209(+13)(-52)$& $3.2\sigma$& 71.8/78& 0.68 \\ \hline
\end{tabular}
\end{ruledtabular}
\label{tab:new_physics_lambda_f3_g2}
\end{table}

Next, we again consider the implications of theoretical uncertainties
in $f_3$ and $g_2$ as relevant for an extraction of $\lambda$ in the
presence of new physics.  
The $\{y_a\}$ parameter set still 
includes $\lambda$ only, while now the $\{y_\mu\}$ parameter set
includes $f_3$ and $g_2$, which are then permitted to vary over a
particular range. 
The resulting 68.3\% confidence levels on $\lambda$ extracted
from a simultaneous fit to the $\{a,A\}$, as well as to the 
individual $\{a\}$ or 
$\{A\}$, New Physics data set, for different theory ranges on
$f_3$ and $g_2$ are compared
in Table \ref{tab:new_physics_lambda_f3_g2}.  It is interesting to
note that the effects of new physics in an $a$ measurement considered
in isolation could potentially be obscured by second class currents, whereas
there is less of an effect for an $A$ measurement considered in isolation or
in a combined fit to $a$ and $A$ data.
This effect
derives from the manner in which $g_2$ appears in $a$. 

\begin{table}
\caption{Fitted values for $\lambda$ and $b_\text{BSM}$ from
  simultaneous, $\{a,A\}$, and individual, $\{a\}$ or $\{A\}$, fits
  to the New Physics data set.  The fitted values for $\lambda$ and
  $b_\text{BSM}$ are defined by the location of the overall
  $\chi^2_\text{min}$, while the 68.3\% CL is defined by
  $\Delta\chi^2 = 2.30$.  We use the inputs $\lambda = 1.2701$ \cite{PDG}
  and $b_\text{BSM} = -0.00522$.  The column ``Prob'' indicates the
  Prob$(\chi^2 > \chi^2_\text{min})$.}
\begin{ruledtabular}
\begin{tabular}{lcccc}
Fit& Fitted $\lambda$& Fitted $b_\text{BSM}$&
  $\chi^2_\text{min}/N_\text{dof}$& Prob \\ \hline
$\{a,A\}$& 1.27011(65)& $-0.0051(31)$& 147.5/156& 0.67 \\
$\{a\}$&   1.27052(113)& $-0.0037(50)$& 73.8/77& 0.58 \\
$\{A\}$&   1.27014(86)& $-0.0045(44)$& 72.2/77& 0.63 \\ \hline
Fit& \multicolumn{4}{c}{Correlation coefficient $\rho_{\lambda b}$} \\ \hline
$\{a,A\}$& \multicolumn{4}{c}{0.98} \\
$\{a\}$&   \multicolumn{4}{c}{0.99} \\
$\{A\}$&   \multicolumn{4}{c}{0.98} \\
\end{tabular}
\end{ruledtabular}
\label{tab:new_physics_lambda_b}
\end{table}

\begin{figure}
\includegraphics[scale=0.42]{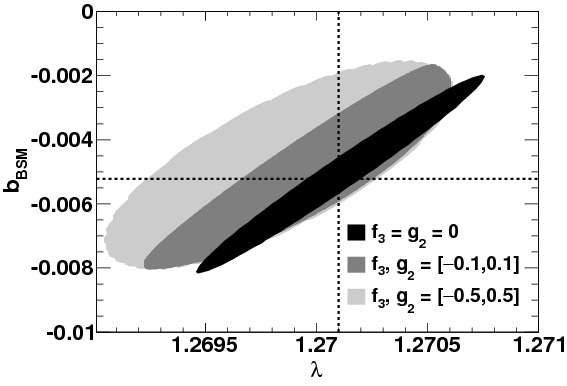} 
\caption{Impact of uncertainties in $f_3$ and $g_2$ on the allowed
  $(\lambda,b_\text{BSM})$ parameter space from a two-parameter
  simultaneous fit to the $\{a,A\}$ New Physics data set.  The bands
  indicate the 68.3\% CL allowed regions defined by $\Delta\chi^2 =
  2.30$ for a joint fit of two free parameters, for the indicated
  uncertainties in $f_3$ and $g_2$.  The input values $\lambda = 1.2701$ and
  $b_\text{BSM} = -0.00522$ are indicated by the dashed lines.}
\label{fig:new_physics_lambda_b_f3_g2}
\end{figure}

Results from the two-parameter fits, with $\{y_a\} = \{\lambda,
b_\text{BSM}\}$ and an empty $\{y_\mu\}$ parameter set, are 
summarized in Table \ref{tab:new_physics_lambda_b}.  It is worth
noting that a combined $\{a,A\}$ fit at this level of precision has
the potential to constrain $b_\text{BSM}$ at 68.3\% CL 
to the level of $\sim 3 \times 10^{-3}$, although such a
  fit would also offer significantly less (by a factor of $\sim 8$) less
  sensitivity to $\lambda$. 
For
completeness, we also list the correlation coefficients
$\rho_{\lambda b}$ for these fits.  Finally, we perform fits with
$\{y_a\} = \{\lambda, b_\text{BSM}\}$ and $\{y_\mu\} = \{f_3,g_2\}$,
with $f_3$ and $g_2$ permitted to vary over different
 particular ranges.  The results
of these fits, which demonstrate the impact that uncertainties in
$f_3$ and $g_2$ have on the allowed $(\lambda, b_\text{BSM})$
parameter space, are shown in
Fig.\ \ref{fig:new_physics_lambda_b_f3_g2}.  Indeed, as one can see, 
the allowed $(\lambda, b_\text{BSM})$
parameter space is broadened significantly by
the inclusion of theoretical uncertainties in $f_3$ and $g_2$.
As per our analysis reported in Table \ref{tab:standard_model_lambda_f3_g2}
we note the uncertainty in $g_2$ is of greater impact. 

\subsection{Example: Impact of the neutron lifetime to an assessment
of the SM}
\label{sec:example_fits_example_tau}

Finally, we consider the impact that future measurements of the
neutron lifetime to a precision of 0.1~s, considered together with
measurements of the $a$ and $A$ angular correlation coefficients to a
precision of $\sim 0.03$\%, will have on an assessment of the validity
of the $V-A$ structure of the SM.  We illustrate this
within the context of our New Physics data set.  To do so, we expand
our $\{x_\text{exp}\}$ data set to $\{a,A,\tau\}$, where we take
$\tau$ to be a single data point whose value is the calculated value
of $\tau = 885.624$~s for the parameters of the New Physics data set, 
as per Table \ref{tab:simulation_input_parameters}, and
consider various uncertainties in $\tau$. 

\begin{table}
\caption{Results for $\chi^2_\text{min}$ from a simultaneous fit for
  $\lambda$ to the $\{a,A,\tau\}$ New Physics data set, for different
  assumed precisions in the measurement of the neutron lifetime, assuming
$g_2=0$.}
\begin{ruledtabular}
\begin{tabular}{ccc}
$\tau_n$ error (s)& $\chi^2_\text{min}/N_\text{dof}$&
  Prob$(\chi^2 > \chi^2_\text{min})$ \\ \hline
1.00 & 155.1/158& 0.55 \\
0.50 & 159.3/158& 0.45 \\
0.25 & 174.3/158& 0.17 \\
0.20 & 183.8/158& 0.078 \\
0.15 & 200.9/158& 0.011 \\
0.10 & 233.0/158& $8.2 \times 10^{-5}$ \\
0.07 & 263.4/158& $2.9 \times 10^{-7}$ \\
\end{tabular}
\end{ruledtabular}
\label{tab:neutron_lifetime_chi_square}
\end{table}

\begin{table}
\caption{Results for $\chi^2_\text{min}$ from a simultaneous fit for
  $\lambda$ and $g_2$ to the $\{a,A,\tau\}$ New Physics data set, for different
  assumed precisions in the measurement of the neutron lifetime, now
for $g_2 \in [-0.025,0]$. 
The first set of values ignore the role of
$g_2$, as per usual procedures,
in the theoretical formula for $\tau$; the second set include it.}
\begin{ruledtabular}
\begin{tabular}{ccc}
$\tau_n$ error (s)& $\chi^2_\text{min}/N_\text{dof}$&
  Prob$(\chi^2 > \chi^2_\text{min})$ \\ \hline
1.00&  154.3/157&  0.55 \\
0.50&  158.0/157&  0.46 \\
0.25&  171.1/157&  0.21 \\
0.20&  179.5/157&  0.11 \\
0.15&  194.5/157&  0.023 \\
0.10&  222.6/157&  $4.5 \times 10^{-4}$ \\
0.07&  249.1/157&  $3.9 \times 10^{-6}$ \\
\hline
1.00&  154.4/157&  0.54 \\
0.50&  158.3/157&  0.46 \\
0.25&  172.1/157&  0.19 \\
0.20&  180.9/157&  0.093 \\
0.15&  196.7/157&  0.017 \\
0.10&  226.3/157&  $2.5 \times 10^{-3}$ \\
0.07&  254.3/157&  $1.4 \times 10^{-6}$ \\
\end{tabular}
\end{ruledtabular}
\label{tab:neutron_lifetime_chi_square_g2s}
\end{table}

\begin{table}
\caption{Results for $\chi^2_\text{min}$ from a simultaneous fit for
  $\lambda$ and $g_2$ to the $\{a,A,\tau\}$ New Physics data set, for different
  assumed precisions in the measurement of the neutron lifetime, now
for $g_2 \in [-0.1,0.1]$. 
The first set of values ignore the role of
$g_2$, as per usual procedures,
in the theoretical formula for $\tau$; the second set include it.}
\begin{ruledtabular}
\begin{tabular}{ccc}
$\tau_n$ error (s)& $\chi^2_\text{min}/N_\text{dof}$&
  Prob$(\chi^2 > \chi^2_\text{min})$ \\ \hline
1.00&  152.7/157&  0.58 \\
0.50&  155.0/157&  0.53 \\
0.25&  163.3/157&  0.35 \\
0.20&  168.6/157&  0.25 \\
0.15&  178.0/157&  0.12 \\
0.10&  195.7/157&  0.019 \\
0.07&  212.5/157&  0.0021 \\
\hline
1.00&  152.9/157&  0.58 \\
0.50&  155.9/157&  0.51 \\
0.25&  166.5/157&  0.29 \\
0.20&  173.3/157&  0.18 \\
0.15&  185.4/157&  0.060 \\
0.10&  208.1/157&  $3.9 \times 10^{-3}$ \\
0.07&  229.6/157&  $1.4 \times 10^{-4}$ \\
\end{tabular}
\end{ruledtabular}
\label{tab:neutron_lifetime_chi_square_g21}
\end{table}

\begin{table}
\caption{Results for $\chi^2_\text{min}$ from a simultaneous fit for
  $\lambda$ and $g_2$ to the $\{a,A,\tau\}$ New Physics data set, for different
  assumed precisions in the measurement of the neutron lifetime, now
for $g_2 \in [-0.5,0.5]$. As in Table \ref{tab:neutron_lifetime_chi_square_g2s}
the first set of values ignore the role of
$g_2$ in the theoretical formula for $\tau$; the second set include it.}
\begin{ruledtabular}
\begin{tabular}{ccc}
$\tau_n$ error (s)& $\chi^2_\text{min}/N_\text{dof}$&
  Prob$(\chi^2 > \chi^2_\text{min})$ \\ \hline
1.00&  152.1/157&  0.60 \\
0.50&  153.0/157&  0.58 \\
0.25&  154.3/157&  0.55 \\
0.20&  154.7/157&  0.54 \\
0.15&  155.0/157&  0.53 \\
0.10&  155.3/157&  0.52 \\
0.07&  155.4/157&  0.52 \\
\hline
1.00&  152.4/157&  0.59 \\
0.50&  154.3/157&  0.55 \\
0.25&  158.6/157&  0.45 \\
0.20&  160.3/157&  0.41 \\
0.15&  162.2/157&  0.37 \\
0.10&  164.3/157&  0.33 \\
0.07&  165.6/157&  0.30 \\
\end{tabular}
\end{ruledtabular}
\label{tab:neutron_lifetime_chi_square_g25}
\end{table}

We then perform a simultaneous fit to the $\{x_\text{exp}\} =
\{a,A,\tau\}$ New Physics data set in which $\{y_a\} = {\lambda}$, 
i.e., with only a single free parameter. We use the CVC value for
$f_2$ and set $f_3=g_2$ to zero. 
As we have already noted, a 
simultaneous fit to the $\{a,A\}$ data set only yields excellent
values for $\chi^2_\text{min}$ --- albeit, with significant pulls on the
fitted values for $\lambda$.  
The neutron lifetime is
relatively insensitive to new scalar or tensor physics, because 
such new
physics only enters quadratically in the $\Xi$ parameter, noting 
Eq.\ (\ref{eq:tau_lambda_Vud}).
Therefore, we would expect
a simultaneous fit to an $\{a,A,\tau\}$ data set in the presence of new
scalar and tensor interactions to return a poor value for 
$\chi^2_\text{min}$.  Indeed, the results of such an analysis are
shown in Table \ref{tab:neutron_lifetime_chi_square}, where we show
the fitted values for $\lambda$, the $\chi^2_\text{min}$ from the fit,
and the probability for that $\chi^2_\text{min}$ value, for different
assumed experimental errors on the lifetime.  As can be seen, 
the probability for the validity of the SM, which we quantify 
via the statistic $\mathcal{P}(\text{SM})
  \leq \text{Prob} (\chi^2 > \chi^2_\text{min})$, as 
 per Sec.\ \ref{sec:likelihood_assessing},  in the
presence of new tensor physics at the level of $g_T \epsilon_T \sim
10^{-3}$ would become $<10^{-4}$ if 
the neutron
lifetime were measured to $\sim 0.1$ s or better 
in concert 
with $\sim 0.03$\%
measurements of $a$ and $A$.  Under this scenario, a precision in
the neutron lifetime of 0.07~s would yield a probability of $\sim 3
\times 10^{-7}$, which is slightly more stringent than the requirement
for a $5\sigma$ result, i.e., a probability of $5.7 \times 10^{-7}$. 

Finally we turn to an examination of the impact of 
a non-zero 
second-class coupling $g_2$ on the ability to falsify 
the $V-A$ law of the SM. It has been the usual procedure to 
ignore certain recoil corrections in the determination of $\tau$, as per 
Eq.~(\ref{eq:tau_lambda_Vud}), 
but we expect that a non-zero value of $g_2$ 
could be important in this context, so that we include the recoil
correction in $g_2$, as per Eq.~(\ref{eq:tau_g2}), as well. 
We  perform a simultaneous fit to the $\{x_\text{exp}\} =
\{a,A,\tau\}$ New Physics data set in which $\{y_a\} = {\lambda}$ 
and $\{y_\mu\} = \{g_2\}$, where $g_2$ is permitted to vary over different
 particular ranges, with $f_2$ equal to its CVC value and 
$f_3=0$. 
The empirical limits on $g_2$ are markedly weaker than 
the existing direct theoretical estimate, 
noting $g_2=-0.0193 \pm 0.0067$ \cite{shiomi96} with 
$\lambda=1.2701$ \cite{PDG}, so that 
we 
perform simultaneous fits using $g_2$ in the
following ranges: $g_2 \in [-0.025,0]$, 
$g_2 \in [-0.1,0.1]$, and 
$g_2 \in [-0.5,0.5]$. 
The fit results, as well as the determined abilities 
to falsify the SM, as a function of the error in the determined
neutron lifetime, are shown  for these ranges of $g_2$ in 
Tables \ref{tab:neutron_lifetime_chi_square_g2s}, 
\ref{tab:neutron_lifetime_chi_square_g21}, 
and 
\ref{tab:neutron_lifetime_chi_square_g25}, respectively. 
Nonzero values of $g_2$ impact the ability to falsify the
SM in every case, and the inclusion of the recoil corrections to $\tau$
are also of importance. 
In the last case, in which 
$g_2 \in [-0.5,0.5]$, the ability to falsify the SM with 
improving precision in the neutron lifetime 
has been completely eroded. Evidently it is important to 
determine $g_2$ to the greatest accuracy possible in order
to be able to falsify the $V-A$ law of the SM.

\section{Summary and Conclusions}
\label{sec:summary}

In summary, we have developed a maximum likelihood statistical
framework, which we term $n$Fitter, 
in which we make simultaneous fits to various neutron $\beta$ decay
observables.  Although a number of global fits to 
$\beta$ decay data have previously been developed \cite{severijns06,konrad10}, 
the novel approach
embedded in our technique is that simultaneous fits to the energy
dependence of the angular correlation coefficients allow for a robust
test of the validity of the $V-A$ structure of the SM in
the presence of theoretical uncertainties, 
whereas fits based on integral quantities do not.  
To our knowledge ours is the first study of the 
quantitative ability to falsify the SM and particularly 
the $V-A$ law, after the manner of 
Refs.~\cite{hocker01,charles05,ciuchini00,ciuchini01,Bona:2005vz,Bona:2006ah,Bona:2007vi}, 
in the context of neutron beta decay observables.

Our study has consisted of fits to the $a$ and $A$ angular
correlation coefficients, as well as to the value of the neutron lifetime. 
We believe that studies of the $B$ angular correlation coefficient, 
as well as of $b_{\rm BSM}$ through the electron energy spectrum in $\beta$ decay,
will offer important complementary information; and such, as well as any
additional, concomitant theoretical uncertainties, can be incorporated
in our analysis framework as well. 
In our current study we have focused
on the role of second-class current contributions, most notably
on the impact of a non-zero  $g_2$ coupling, on the ability to identify 
physics BSM.  
In the course of developing our analysis procedure, 
we have discovered that certain recoil effects to the neutron lifetime, 
contrary to the 
usual view \cite{wilkinson82}, 
can have an impact on our fit results. Moreover, the precise
form of the recoil corrections depends on experimental details, 
revealing that the corrections change 
with a finite
experimental acceptance, such as in experiments which extract a value
of the lifetime from measurements of the decay electrons and/or
protons. Such considerations warrant further detailed study.

We have explicitly shown that it is possible to 
{\it discover} physics BSM, at $5\sigma$ significance, 
 in neutron $\beta$ decay observables using
experiments which are {\it currently} planned or under construction. 
This is subject to the following conditions; namely, that 
(i) tensor interactions are not much smaller than the constraint 
which emerges from the Dalitz analysis of pion radiative $\beta$ 
decay \cite{Bychkov:2008ws,Cirigliano:2013xha}, (ii) the value of
$g_2$ can be sharply restricted, and (iii) results of 
0.03\% precision can be realized for $a$ and $A$, in concert 
with a sub-$0.1\, {\rm s}$ determination of the neutron lifetime. 
In our study we have assumed that $g_T {\rm Re}(\epsilon_T)=0.001$, so that
using $g_T=1.05 (35)$ \cite{bhattacharya12} and noting 
${\rm Re}(\epsilon_T) \sim v^2/\Lambda_{\rm BSM}^2$, 
this would be commensurate with the appearance of physics BSM at an energy scale 
of at least $\Lambda_{\rm BSM} \sim 5\, {\rm TeV}$ \cite{Appelquist:1974tg}. 
We note that existing direct limits on tensor couplings
from nuclear $\beta$ decay are much weaker than those from radiative pion 
decay \cite{Cirigliano:2013xha}; perhaps new physics effects could be different
in pion and neutron decays. We note, 
however, that under the assumption that BSM effects appear at energies in excess of 
$\Lambda_{\rm BSM}$ 
such effects can only occur from 
operators beyond mass-dimension six and ought be suppressed. 
Our current analysis framework is also suitable to 
the discovery of new scalar interactions as well. 

We have shown that theoretical uncertainties in $g_2$ 
can 
mitigate the gains made in falsifying the SM through the
inclusion of 
precision $\tau$ results. We thus advocate for a determination of
$g_2$ using lattice gauge theory techniques; we suppose that 
lattice measurements of $f_2$ and $f_3$ in neutron decay would be useful, too. 
These considerations are 
quite independent of how information on $b_{\rm BSM}$ is determined. 

\begin{acknowledgments}

S.G.\ is supported in part by the Department of Energy Office of
Nuclear Physics under Grant No.\ DE-FG02-96ER40989.
B.P.\ is supported in part by the Department of Energy Office of
Nuclear Physics under Grant No.\ DE-FG02-08ER41557.

\end{acknowledgments}


\end{document}